\documentclass[sigconf]{acmart}
\usepackage{balance}
\usepackage{makecell}
\usepackage{subfigure}
\renewcommand\footnotetextcopyrightpermission[1]{}

\AtBeginDocument{%
  \providecommand\BibTeX{{%
    \normalfont B\kern-0.5em{\scshape i\kern-0.25em b}\kern-0.8em\TeX}}}

\begin{document}

\title{Semi-Disentangled Representation Learning in Recommendation System}

\author{Weiguang Chen}
\email{cwg@hnu.edu.cn}
\affiliation{%
	\institution{Hunan University}
}

\author{Wenjun Jiang}
\email{jiangwenjun@hnu.edu.cn}
\authornote{Wenjun Jiang is the corresponding author.}
\affiliation{%
	\institution{Hunan University}
}

\author{Xueqi Li}
\email{lee\_xq@hnu.edu.cn}
\affiliation{%
	\institution{ Hunan University}
}


\author{Kenli Li}
\email{lkl@hnu.edu.cn}
\affiliation{%
	\institution{Hunan University}
}

\author{Albert Zomaya}
\email{albert.zomaya@sydney.edu.au}
\affiliation{%
	\institution{University of Sydney}
}

\author{Guojun Wang}
\email{csgjwang@gzhu.edu.cn	}
\affiliation{%
	\institution{Guangzhou University}
}

\begin{abstract}
	Disentangled representation has been widely explored in many fields due to its maximal compactness, interpretability and versatility. Recommendation system also needs disentanglement to make representation more explainable and general for downstream tasks. However, some challenges slow its broader application --- the lack of fine-grained labels and the complexity of user-item interactions. To alleviate these problems, we propose a Semi-Disentangled Representation Learning method (SDRL) based on autoencoders. SDRL divides each user/item embedding into two parts: the explainable and the unexplainable, so as to improve proper disentanglement while preserving complex information in representation. The explainable part consists of $internal\ block$ for individual-based features and $external\ block$ for interaction-based features. The unexplainable part is composed by $other\ block$ for other remaining information. Experimental results on three real-world datasets demonstrate that the proposed SDRL could not only effectively express user and item features but also improve the explainability and generality compared with existing representation methods.
\end{abstract}

\maketitle

\section{Introduction}
Disentangled representation learning aims at separating embedding into multiple independent parts, which makes it more explainable and general \cite{bengio_representation_2013}. The core idea of existing methods is to minimize the reconstruction error of the whole representation and maximize the independence among different parts simultaneously \cite{chen_infogan_2016,higgins_beta-vae_2017}. It has been successfully applied into image representation, and researchers have verified its superiority on many downstream tasks, e.g., image generation \cite{alharbi_disentangled_2020,chen_infogan_2016} and style transfer \cite{kotovenko_content_2019}. Disentangled representation is also required by recommendation system to distinguish various hidden intentions under the same behavior \cite{wang_disentangled_2020,ma_disentangled_2020}. However, two obstacles slow its extensive application: the lack of enough fine-grained labels and the complexity of user-item interactions. Furthermore, Locatello et al. \cite{locatello_challenging_2019} theoretically demonstrate the difficulty and even impossibility of unsupervised disentanglement and propose solutions using a few labels. It inspires us to put forward a Semi-Disentangled Representation Learning (SDRL) approach for recommendation system based on limited labels.

\begin{figure}
	\centering
	\includegraphics[width=0.85\linewidth] {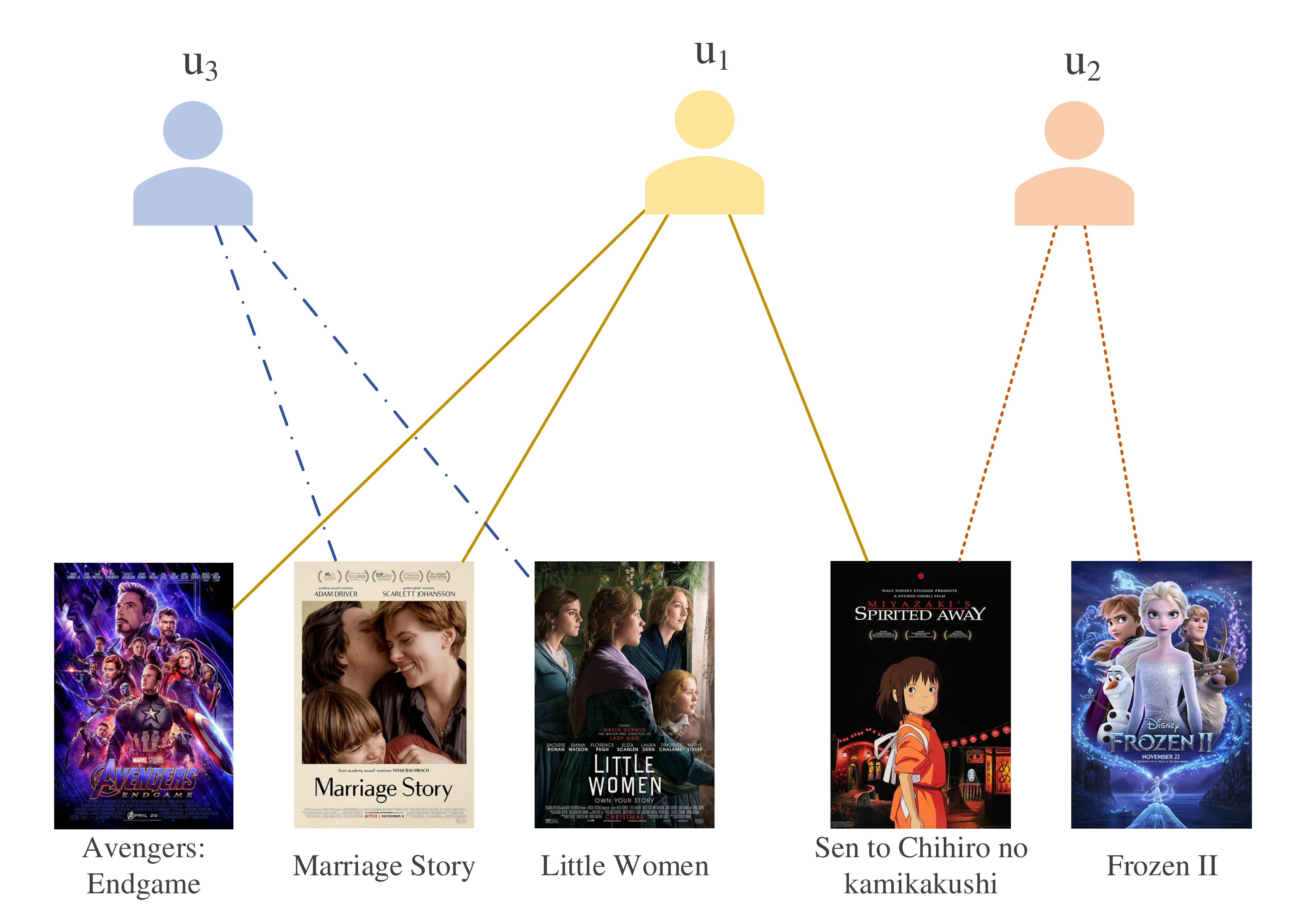}
	\vspace{-1em}
	\caption{An Illustration of User-Item Interactions \protect\footnotemark[1].}
	\label{fig:example}
\end{figure}

\footnotetext[1]{Images come from https://www.imdb.com/.}

Specifically, we would introduce an example with Fig. \ref{fig:example} to explain the requirements and challenges of disentangled representation for recommendation and clarify our motivations in this paper. For instance, users $u_1$ and $u_3$ have watched the same movie {\em Marriage Story}, but their motivations may differ from each other. $u_1$ chooses it probably because he is a fan of the actress Scarlett Johansson while the motivation of $u_3$ maybe that he takes interests in romance movies. Another scene (i.e., users $u_1$ and $u_2$ watch the same movie {\em Sen to Chihiro no kamikakushi}) reflects a similar phenomenon. $u_2$ likes animation movies while $u_1$ also watches it perhaps because $u_1$ and $u_2$ are friends. Disentangled representation could help distinguish these different intentions under the same behavior, so as to improve the representation accuracy and offer clues to explain why the items are provided.

However, it is hardly possible to develop complete and accurate disentanglement in recommendation system, in consideration that it lacks fine-grained labels and user-item interactions are complicated. Specific with the example, it usually lacks enough labels for building completely fine-grained aspects, e.g., $u_1$ is a fan of Scarlett Johansson.  On the other hand, some unknowable and random factors also affect users' decisions, e.g., $u_2$ invites his friend $u_1$ to watch {\em Sen to Chihiro no kamikakushi} together. Disentanglement based on incomplete or inaccurate factors may decrease the expression ability of representation.

To overcome these limitations, we propose a semi-disentangled representation method SDRL, separating representation into three parts to respectively express internal, external and other remaining complex information. In particular, we present $internal\ block$ to denote the features related to individual itself, e.g., product category, movie style, user age. $external\ block$ represents the characteristics from user-item interactions, e.g., user ratings and implicit feedback. Besides, we introduce the $other\ block$ to generalize the information that may not contained by the former two blocks or random factors in the real scenes. Moreover, in addition to reduce  the overall reconstruction error, we utilize category information and user-item ratings as the supervision for $internal\ block$ and $external\ block$ respectively. In this way, SDRL could not only capture complex interactive information but also express various features into different blocks. To sum up, the main contributions are as follows:

\begin{itemize}
	\item We identify the problem of semi-disentangled representation learning in recommendation system, to preserve complex information while achieving proper disentanglement. As far as we know, we are the first to study this problem.
	\item We propose a method SDRL to address the problem. It divides the representation into three blocks: $internal\ block$, $external\ block$ and $other\ block$. The former two are employed to express individual- and interaction-based features. The $other\ block$ contains remaining information.
	\item The experimental results demonstrate that the proposed SDRL improves the accuracy of two downstream tasks, as well as enhances the explainablity of representation.
\end{itemize}

\section{Task Formulation}
To improve both disentanglement and expression ability of representation in recommendation system, we separate embeddings into three blocks: $internal\ block$, $external\ block$ and $other\ block$. We would formally define these concepts and the problem to solve in this paper.

\subsection{Key Concepts}

{\bf Definition 1: internal block.} It contains features about individual itself, which are extracted from content information. 

{\bf Definition 2: external block.} It expresses features based on interactions, i.e., user-item ratings, implicit feedback, etc. 

{\bf Definition 3: other block.} $other\ block$ denotes characteristics excluding those contained by the former blocks.

In this work, we utilize category information to supervise $internal\ block$. User-item ratings are employed as the supervision of $external\ block$. $other\ block$ generalizes some features belonged to individuals and interactions but beyond supervision (e.g., the color of a product or the director of a movie), as well as some random factors.

\subsection{Problem Definition}
We identify the problem of semi-disentangled representation in recommendation system, to preserve complicated information while strengthening the representation explainability and generality with proper disentanglement. It requires to embed all features into three blocks using limited labels as supervision. It could be formally defined as follows.

{\bf Input:} The normalized user-item ratings $R$ and item categories $C^I$ are taken as the input, in which $R_{i,j}=r$ denotes that the $i$-th user rates $j$-th item as $r$ points and $C^I_{p,q}=1$ means that the $p$-th item belongs to $q$-th category. 

{\bf Output:} Each user and item needs to be represented as a $k$-dimension vector consisting of three blocks,  $internal\ block$, $external\ block$ and $other\ block$.

{\bf Objective:} The goal is to (1) make representation $Z$ preserve more original information of users and items and (2) encourage $internal\ block$ and $external\ block$ to express more features related to categories and interactions respectively. The objective function is defined as follows.

\begin{gather}
Maximize\ P(U,I,R|Z)+P(C|Z_{int})+P(R|Z_{ext}),
\end{gather}
where $U$ and $I$ represent initial embeddings of users and items, $Z$ denotes their representation in semi-disentangled space, $Z_{int}$ and $Z_{ext}$ are the corresponding $internal\ block$ and $external\ block$, and $C=C^I\cup C^U$ contains category labels of items and users. $P(A|B)$ means the possibility of generating $A$ given $B$.

\begin{figure}[h]
	\centering
	\includegraphics[width=\linewidth] {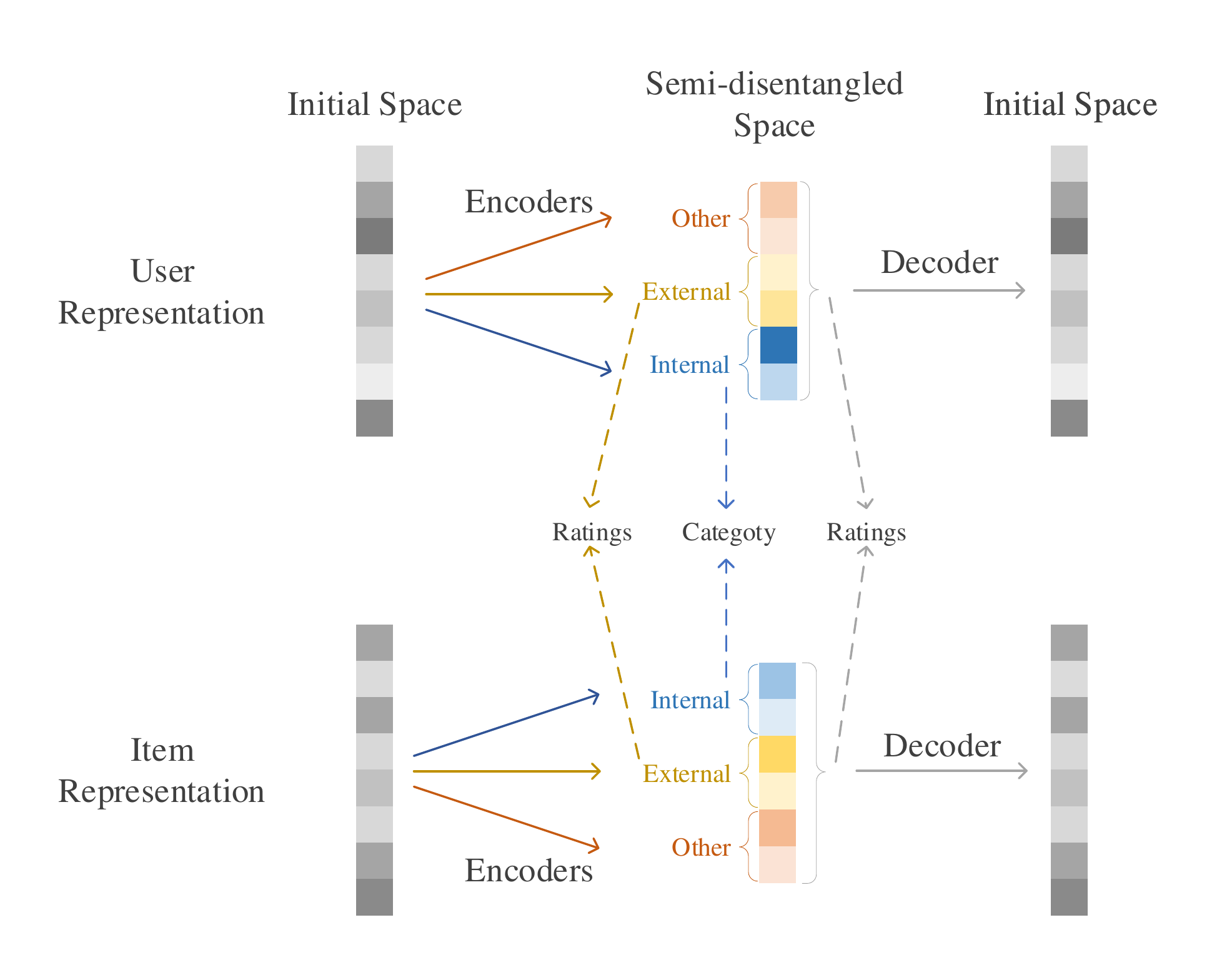}
	\vspace{-2em}
	\caption{The Framework of SDRL.}
	\label{fig:framework}
\end{figure}

\section{SDRL: The Details} \label{sec:method}
We propose a semi-disentangled representation learning method SDRL for recommendation system. The framework is shown in Fig. \ref{fig:framework}. It consists of two major components. (1) {\em Node representation} (i.e., user representation and item representation) employs autoencoders to preserve characteristics of users, items and their interactions. (2) {\em Supervised semi-disentanglement} utilizes category information and user-item ratings to encourage $internal\ block$ and $external\ block$ to respectively express more individual and interactive features. We also point out some possible directions to extend SDRL.

\subsection{Node Representation}
We exploit autoencoders to transform representation in initial space into semi-disentangled space. 

Autoencoder is an unsupervised deep neural network \cite{hinton_autoencoders_1993,vincent_stacked_2010}, which has been extensively applied for network representation learning \cite{wu_collaborative_2016,tallapally_user_2018}. It has two modules of encoder and decoder as follows,

\begin{gather}
f(x)=\sigma(Wx+b), \\
g(y)=\sigma(W'y+b').
\end{gather}
Encoder denotes the mapping $f$ that transforms the input vector $x\in R^d$ into the latent representation $y\in R^{d'}$, where $W \in R^{d'\times d}$ is a weight matrix and $b \in R^{d'}$ is an offset vector. The mapping $g$ is called the decoder, reconstructing $y$ in the latent space into $x'$ in the initial space, in which $W' \in R^{d\times d'}$ and $b' \in R^{d}$ denote the weight matrix and the offset vector respectively. $\sigma(.)$ is an activation function. 

The objective of autoencoder is to minimize the reconstruction error. Stacked autoencoders \cite{vincent_stacked_2010} is a widely used variant, which has been experimentally verified the improvement of representation quality. Therefore, we employ it to generate representations of users and items.

Different from typical autoencoders, we use three encoders $Encoder_{int}$, $Encoder_{ext}$ and $Encoder_{oth}$ to produce the corresponding $internal\ block$, $external\ block$ and $other\ block$, respectively. The generation process of users is as follows, 

\begin{gather}
Z^U_{int}=Encoder_{int}(U), \\
Z^U_{ext}=Encoder_{ext}(U), \\
Z^U_{oth}=Encoder_{oth}(U), \\
Z^U=concentrate(Z_{int},Z_{ext},Z_{oth}), \\
\hat{U}=Decoder(Z^U). 
\end{gather}
$Z^U_{int}$, $Z^U_{ext}$ and $Z^U_{oth}$ represent the above three blocks respectively, $Z^U$ denotes the embeddings of users $U$ in semi-disentangled space, and $\hat{U}$ is the generated representations of users in the initial space. The representation of items also utilizes a similar process. The goal of the process is to reconstruct the representation of users and items in the initial space as well as user-item ratings with autoencoders. Considering that the number of unobserved interactions (i.e., there is no rating) far exceeds that of the observed, we employ Binary Cross Entropy (BCE) as the basic metric. We define the loss function of reconstruction as follows,

\begin{gather}
loss_{recon}=BCE(\hat{U},U)+BCE(\hat{I},I)+BCE(\hat{R},R), \label{eq:recon}
\end{gather}
where $\hat{U}$, $\hat{I}$ denote the reconstructed users and items, $\hat{R}$ represent predicted ratings by matching $Z^U$ and $Z^I$. Details of BCE are as follow, where  $y$ denote labels, $\hat{y}$ represent the predicted values and $N$ is the number of $y$, 

\begin{gather}
BCE(\hat{y},y)=\frac{1}{N}\sum_{i=1}^{N}y_i\log(\hat{y_i})+(1-y_i)\log(1-\hat{y_i}).
\end{gather}

\subsection{Supervised Semi-Disentanglement}
Disentangled representation with weakly supervision has demonstrated its effectiveness in computer vision \cite{locatello_disentangling_2020,chen_weakly_2020}. The successful application and the complexity of interactions inspires us to improve the representation disentanglement using limited labels in recommendation system. 

\subsubsection{Internal Block Supervision}
We employ category information $C$ as the supervision for $internal\ block$. $C^I$ represents the item-category correspondence based on side information. $C^U$ denotes the user preference on category extracted from ratings and $C^I$, which is calculated as follows,

\begin{gather}
C^{U_m}=Normalize(\sum_{I_n \in I, R_{m,n}>0}R_{m,n}C^{I_n}).
\end{gather}
$C^{U_m}$ and $C^{I_n}$ respectively denote category vectors of $U_m$ and $I_n$. We sum the product of rating $R_{m,n}$ and item category vector $C^{I_n}$ of all rated items, and normalize it as $U_m$'s category preferences $C^{U_m}$. The loss function is as follows, 

\begin{gather}
loss_{int}=BCE(\hat{C_{int}},C)), \label{eq:int}
\end{gather}
where $\hat{C_{int}}$ denotes the predicted category features of users and items using the corresponding $internal\ block$.

\subsubsection{External Block Supervision}
We utilize ratings $R$ for supervising $external\ block$ to contain more interactive information. The loss function on $external\ block$ is as follows,

\begin{gather}
loss_{ext}=BCE(\hat{R_{ext}},R)). \label{eq:ext}
\end{gather}
$\hat{R_{ext}}$ denote the predicted ratings  based on $external\ block$. 

\subsubsection{Semi-Disentanglement Analysis}
Most existing methods explicitly improve the block independence with the mutual information or distance correlation \cite{cheng_improving_2020}, which maybe not well applicable in recommendation. The major reason is that we need improve representation disentanglement as well as preserve interrelated characteristics. Hence, we propose semi-disentangled method SDRL. It does not separate the whole embedding into explainable blocks as that in disentangled representation learning, i.e., preserving no explicit meanings or factors in $other\ block$. Furthermore, it does not force the independence among different blocks. Instead, it just pushes the explainable blocks to express the more corresponding characteristics, i.e., encouraging $internal\ block$ to express more category-based information and $external\ block$ to contain more interactive features.

\subsection{Model Optimization and Extension}
We aim at improving the expression ability and proper disentanglement of representation at the same time, so we combine the loss function of node representation and semi-disentanglement as follows,

\begin{gather}
loss=loss_{recon}+loss_{int}+loss_{ext}. \label{eq:all}
\end{gather}

\subsubsection{Setting-oriented Extension}
We separate the explainable part of representation into two blocks using autoencoders. As data characteristics and label information change, it's flexible to adjust the number and the type of blocks as well as switch basic representation method. There are some possible extensions.

When more fine-grained labels are available even not for each sample, it is reasonable and convenient to add more blocks and optimize them with a small number of labels as in  \cite{locatello_disentangling_2020}. In another setting where the extracted features are independent from each other, variational autoencoders (VAE) \cite{kingma_introduction_2019} maybe a good choice to replace autoencoders.

\subsubsection{Task-oriented Extension}
The objective of this work is to produce general representation, so just a simple match of embeddings is employed for various downstream tasks. In fact, different tasks may emphasize different factors. For a specific task, some following-up modules maybe required and it is easy to extend SDRL with them.

For instance, for some tasks with supervision (e.g., rating prediction, node classification), attention mechanism \cite{vaswani_attention_2017} is an intuitively excellent option as the following-up module to allocate different weights for different factors. For some tasks without labels (e.g., serendipity recommendation \cite{li_directional_2020}), pre-assigned weights probably could improve the performance. In brief, the proposed semi-disentangled representation provides opportunity to adaptively or manually pay different attentions to known factors on various tasks. 

\section{Experiments}
To validate the effectiveness and explainability of SDRL, as well as the role of various semi-disentangled blocks, we conduct intensive experiments on three real-world datasets. 
We would briefly introduce the experimental settings in Section \ref{sec:setting}. Section \ref{sec:comparison} displays the comparison among our method and baselines in Top-K recommendation and item classification. In Sections \ref{sec:ablation} and \ref{sec:para}, we perform ablation experiments and hyper-parameter analysis to study the impacts of different blocks and parameters in SDRL. Finally, we demonstrate the semi-disentanglement and explainability of representation with visualization and a case study in Section \ref{sec:vis}.

\subsection{Experimental Settings} \label{sec:setting}
We perform experiments on three real-world datasets: MovieLens-latest-small (ml-ls),  MovieLens-10m (ml-10m) \cite{harper_movielens_2016} and Amazon-Book \cite{mcauley_image-based_2015,he_ups_2016}, whose statistics are shown in Table \ref{tab:dataset}. We filter out users and items with less than 20 ratings and employ 80\% ratings as the training data and the others as test data. We compare our method with four baselines, NetMF \cite{qiu_network_2018,rozemberczki_karate_2020}, ProNE \cite{zhang_prone_2019}, VAECF \cite{liang_variational_2018} and MacridVAE \cite{ma_learning_2019}, based on four metrics, $recall$, $precision$, $F1$ and $ndcg$ \cite{shani_evaluating_2011}. 


\begin{table}[h]
	\centering
	\small
	\caption{Statistics of Datasets.}
	\vspace{-1em}
	\begin{tabular}{c|ccccc}
		\hline
		dataset&\#Users&\#Items&\#Ratings&\#Categories&Density \\ \Xhline{1.5pt}
		ml-ls & 610&9742&100836&18&1.7\% \\
		ml-10m & 71567 &10681 &10000054 &18&1.3\% \\
		Amazon-Book & 52643&91599&2984108&22&0.062\% \\ \hline
	\end{tabular}
	\label{tab:dataset}
\end{table}

\subsubsection{Baselines} 
We would briefly introduce four baselines and four variants of our proposed method SDRL.

{\bf NetMF:} NetMF \cite{qiu_network_2018} makes network embedding as matrix factorization, unifying four network embedding methods DeepWalk \cite{perozzi_deepwalk_2014}, LINE \cite{tang_line_2015}, PTE \cite{tang_pte_2015}, and node2vec \cite{grover_node2vec_2016}.


{\bf ProNE:} ProNE \cite{zhang_prone_2019} is a fast and scalable network representation approach consisting of two modules, sparse matrix factorization and propagation in the spectral space.

{\bf VAECF:} Liang et al. develop a variant of Variational AutoEncoders for Collaborative Filtering \cite{liang_variational_2018} on implicit feedback data, which is a non-linear probabilistic model. 

{\bf MacridVAE:} MacridVAE \cite{ma_learning_2019} is one of the state-of-the-art methods that learn disentangled representation for recommendation, achieving macro and micro disentanglement.

To study the impacts of three blocks, $internal\ block$ (int), $external\ block$ (ext) and $other\ block$ (oth), we develop some variants through different combinations.  For variants with two blocks, we set their proportion as 1:1.

{\bf SDRL'(int+ext)} generates node embeddings consisting of two blocks, $internal\ block$ and $external\ block$.

{\bf SDRL'(int+oth)} keeps $internal\ block$ and $other\ block$.    

{\bf SDRL'(ext+oth)} consists of $external\ block$ and $other\ block$.

{\bf SDRL'(whole)} represents a node as a whole embedding, which is similar to those common representation methods, but its optimization is based on autoencoders as in SDRL. 

\subsection{Performance Comparison} \label{sec:comparison}
We verify the effectiveness of SDRL in two common downstream tasks in recommendation system, Top-K recommendation and item classification. We run the experiments 20 times and report the average values and the standard deviation. We highlight the best values of baselines in bold and calculate the corresponding improvements. 
\begin{table*}[!ht]
	\centering
	\caption{Comparison on Top-K recommendation(\%). (ml-ls)}
	\vspace{-1em}
	\begin{tabular}{c|ccc|ccc}
		\hline
		method&F1@5&F1@10&F1@15&ndcg@5&ndcg@10&ndcg@15\\ \Xhline{1.5pt}
		NetMF&2.9573($\pm$0.0)&4.7776($\pm$0.0)&5.1616($\pm$0.0)&3.793($\pm$0.0)&5.2728($\pm$0.0)&6.9132($\pm$0.0)\\
		ProNE&3.0491($\pm$0.1)&4.5284($\pm$0.14)&5.3116($\pm$0.17)&3.8125($\pm$0.18)&5.243($\pm$0.13)&7.0057($\pm$0.11)\\
		VAECF&\bf{5.5595($\pm$0.29)}&\bf{7.7409($\pm$0.33)}&\bf{8.9691($\pm$0.43)}&\bf{8.7243($\pm$0.55)}&\bf{9.8197($\pm$0.36)}&\bf{11.3705($\pm$0.35)}\\
		MacridVAE&5.1578($\pm$0.15)&7.3867($\pm$0.26)&8.5167($\pm$0.32)&8.0452($\pm$0.33)&9.0503($\pm$0.32)&10.6263($\pm$0.23)\\
		\hline
		improvement&30.0171&27.9192&25.4228&20.4062&25.8338&28.8774\\
		\hline
		SDRL&7.2283($\pm$0.27)&9.9021($\pm$0.27)&11.2493($\pm$0.26)&10.5046($\pm$0.57)&12.3565($\pm$0.3)&14.654($\pm$0.26)\\
		\hline
	\end{tabular}
	\label{tab:topk0}
\end{table*}

\begin{table*}[!ht]
	\centering
	\caption{Comparison on Top-K  recommendation(\%). (ml-10m)}
	\vspace{-1em}
	\begin{tabular}{c|ccc|ccc}
		\hline
		method&F1@5&F1@10&F1@15&ndcg@5&ndcg@10&ndcg@15\\ \Xhline{1.5pt}
		NetMF&3.9352($\pm$0.0)&6.2($\pm$0.0)&7.4752($\pm$0.0)&4.831($\pm$0.0)&6.8198($\pm$0.0)&9.2123($\pm$0.0)\\
		ProNE&2.1065($\pm$0.04)&3.1941($\pm$0.06)&3.8371($\pm$0.05)&2.3357($\pm$0.05)&3.5148($\pm$0.04)&4.9629($\pm$0.03)\\
		VAECF&5.1655($\pm$0.07)&\bf{7.5696($\pm$0.1)}&\bf{8.881($\pm$0.1)}&\bf{8.033($\pm$0.18)}&\bf{9.4863($\pm$0.13)}&\bf{11.4674($\pm$0.11)}\\
		MacridVAE&\bf{5.1832($\pm$0.07)}&7.5236($\pm$0.11)&8.8513($\pm$0.14)&7.0934($\pm$0.18)&9.0107($\pm$0.16)&11.4481($\pm$0.15)\\
		\hline
		improvement&44.6963&38.8184&35.2393&42.2793&38.2288&37.0189\\
		\hline
		SDRL&7.4999($\pm$0.16)&10.508($\pm$0.23)&12.0106($\pm$0.25)&11.4293($\pm$0.46)&13.1128($\pm$0.35)&{15.7125($\pm$0.27)}\\
		\hline
	\end{tabular}
	\label{tab:topk1}
\end{table*}

\begin{table*}[!ht]
	\centering	
	\caption{Comparison on Top-K recommendation(\%). (Amazon-Book)}
	\vspace{-1em}
	\begin{tabular}{c|ccc|ccc}
		\hline
		method&F1@5&F1@10&F1@15&ndcg@5&ndcg@10&ndcg@15\\ \Xhline{1.5pt}
		NetMF&\bf{9.0322($\pm$0.0)}&\bf{10.9837($\pm$0.0)}&\bf{11.5754($\pm$0.0)}&\bf{13.4461($\pm$0.0)}&\bf{12.623($\pm$0.0)}&\bf{13.8484($\pm$0.0)}\\
		ProNE&6.818($\pm$0.07)&8.8102($\pm$0.05)&9.5218($\pm$0.07)&9.9492($\pm$0.14)&9.9032($\pm$0.06)&11.1619($\pm$0.04)\\
		VAECF&6.1366($\pm$0.09)&7.9531($\pm$0.14)&8.6195($\pm$0.14)&9.6312($\pm$0.19)&9.3344($\pm$0.14)&10.3365($\pm$0.1)\\
		MacridVAE&8.0043($\pm$0.12)&9.9379($\pm$0.14)&10.5185($\pm$0.17)&12.1028($\pm$0.22)&11.5303($\pm$0.13)&12.6398($\pm$0.11)\\
		\hline
		improvement&
		0.2879&4.3592&5.9549&2.0549&5.4543&7.1127\\
		\hline
		SDRL&{9.0582($\pm$0.07)}&11.4625($\pm$0.12)&12.2647($\pm$0.14)&{13.7224($\pm$0.18)}&13.3115($\pm$0.14)&14.8334($\pm$0.1)\\
		\hline
	\end{tabular}
	\label{tab:topk2}
\end{table*}

\begin{figure}[h]
	\centering
	\subfigure[ml-ls]{
		\includegraphics[width=0.3\linewidth]{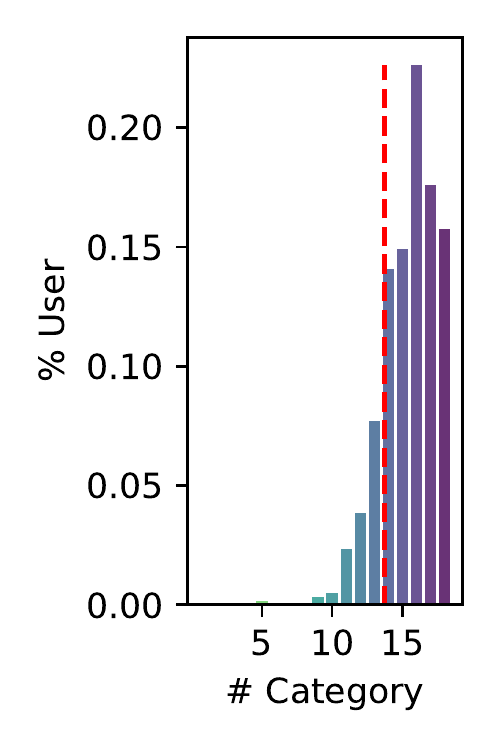} 
		\label{fig:sta_gen0}
	}
	\subfigure[ml-10m]{
		\includegraphics[width=0.3\linewidth]{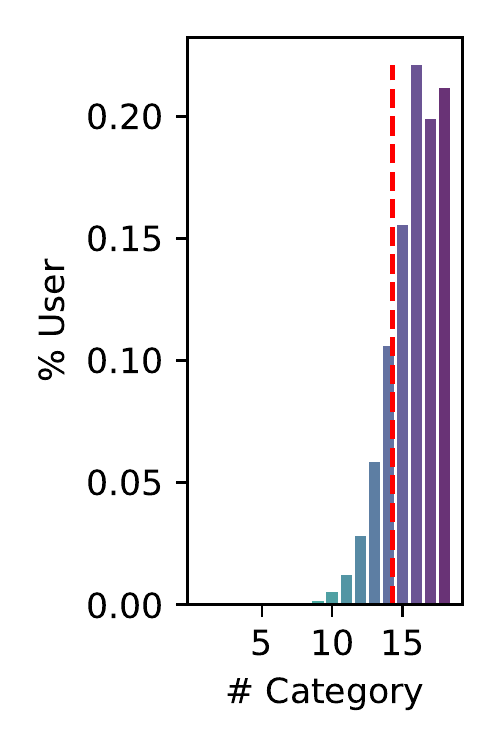} 
		\label{fig:sta_gen1}
	}
	\subfigure[Amazon-Book]{
		\includegraphics[width=0.3\linewidth]{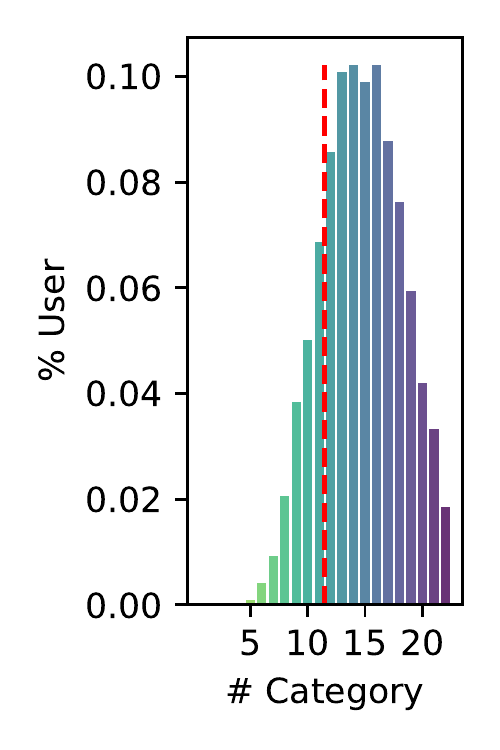} 
		\label{fig:sta_gen2}
	}
	\vspace{-1em}
	\caption{Statistics of Categories Grouped by Users.}
	\label{fig:sta_gen}
\end{figure}

\begin{table}[ht]
	\centering
	\small
	\caption{Comparison on item classification(\%). (ml-ls)}
	\vspace{-1em}
	\begin{tabular}{c|ccc}
		\hline
		method&recall&precison&micro\_F1\\ \Xhline{1.5pt}
		NetMF&50.3539&59.8585&54.6957\\
		ProNE&\bf{53.1639}&\bf{62.73}&\bf{57.5517}\\
		\hline
		improvement&12.4686&13.6479&13.0066\\
		\hline
		SDRL&59.7927&71.2913&65.0372\\
		\hline
	\end{tabular}
	\label{tab:cla0}
\end{table}

\begin{table}[ht]
	\centering
	\small
	\caption{Comparison on item classification(\%). (ml-10m)}
	\vspace{-1em}
	\begin{tabular}{c|ccc}
		\hline
		method&recall&precison&micro\_F1\\ \Xhline{1.5pt}
		NetMF&66.4181&65.6991&66.0564\\
		ProNE&\bf{67.7764}&\bf{66.7775}&\bf{67.2731}\\
		\hline
		improvement&18.0496&19.9254&18.9872\\
		\hline
		SDRL&{80.0098}&{80.0832}&{80.0464}\\
		\hline
	\end{tabular}
	\label{tab:cla1}
\end{table}

\begin{table}[ht]
	\centering
	\small
	\caption{Comparison on item classification(\%). (Amazon-Book)}
	\vspace{-1em}
	\begin{tabular}{c|ccc}
		\hline
		method&recall&precison&micro\_F1\\ \Xhline{1.5pt}
		NetMF&\bf{51.4335}&\bf{47.3786}&\bf{49.3226}\\
		ProNE&51.2197&47.3555&49.2115\\
		\hline
		improvement&83.7268&76.2521&79.759\\
		\hline
		SDRL&94.4971&83.5058&88.6618\\
		\hline
	\end{tabular}
	\label{tab:cla2}
\end{table}

\begin{table*}[!ht]
	\centering
	\caption{Comparison on Top-K recommendation with different combinations of blocks (\%). (ml-ls)}
	\vspace{-1em}
	\begin{tabular}{c|ccc|ccc}
		\hline
		method&F1@5&F1@10&F1@15&ndcg@5&ndcg@10&ndcg@15\\ \Xhline{1.5pt}
		SDRL&\bf{7.2283($\pm$0.27)}&\underline{9.9021($\pm$0.27)}&\underline{11.2493($\pm$0.26)}&\underline{10.5046($\pm$0.57)}&\underline{12.3565($\pm$0.3)}&\underline{14.654($\pm$0.26)}\\
		SDRL'(int+ext)&\underline{6.9607($\pm$0.22)}&\underline{9.7205($\pm$0.24)}&\underline{11.1285($\pm$0.27)}&9.8062($\pm$0.48)&12.0343($\pm$0.31)&\underline{14.5021($\pm$0.25)}\\
		SDRL'(int+oth)&\underline{7.1316($\pm$0.21)}&\bf{9.9035($\pm$0.26)}&\bf{11.3039($\pm$0.24)}&\underline{10.247($\pm$0.51)}&\underline{12.2864($\pm$0.29)}&\bf{14.8077($\pm$0.19)}\\
		SDRL'(ext+oth)&6.7585($\pm$0.28)&9.2792($\pm$0.19)&10.5929($\pm$0.2)&9.5182($\pm$0.55)&11.5124($\pm$0.26)&13.985($\pm$0.18)\\
		SDRL'(whole)&6.2179($\pm$0.75)&9.146($\pm$0.59)&10.4072($\pm$0.51)&\bf{12.796($\pm$1.96)}&\bf{13.1785($\pm$1.05)}&13.864($\pm$0.6)\\
		\hline
	\end{tabular}
	\label{tab:topk0a}
\end{table*}

\begin{table*}[!ht]
	\centering
	\caption{Comparison on Top-K  recommendation with different combinations of blocks (\%). (ml-10m)}
	\vspace{-1em}
	\begin{tabular}{c|ccc|ccc}
		\hline
		method&F1@5&F1@10&F1@15&ndcg@5&ndcg@10&ndcg@15\\ \Xhline{1.5pt}
		SDRL&\bf{7.4999($\pm$0.16)}&\bf{10.508($\pm$0.23)}&\bf{12.0106($\pm$0.25)}&\underline{11.4293($\pm$0.46)}&\underline{13.1128($\pm$0.35)}&\bf{15.7125($\pm$0.27)}\\
		SDRL'(int+ext)&\underline{6.8778($\pm$0.23)}&\underline{9.5076($\pm$0.34)}&\underline{10.8903($\pm$0.44)}&\underline{10.5319($\pm$1.02)}&\underline{11.8883($\pm$0.81)}&\underline{14.2037($\pm$0.75)}\\
		SDRL'(int+oth)&\underline{7.432($\pm$0.29)}&\underline{10.4417($\pm$0.38)}&\underline{11.9863($\pm$0.44)}&\bf{11.936($\pm$1.22)}&\bf{13.2881($\pm$0.88)}&\underline{15.5704($\pm$0.74)}\\
		SDRL'(ext+oth)&5.7576($\pm$0.22)&8.4256($\pm$0.28)&9.886($\pm$0.46)&8.5747($\pm$0.77)&10.3502($\pm$0.62)&12.8529($\pm$0.49)\\
		SDRL'(whole)&5.1166($\pm$0.39)&7.5039($\pm$0.37)&8.7737($\pm$0.53)&10.1397($\pm$2.41)&10.757($\pm$2.44)&11.8431($\pm$2.08)\\
		\hline
	\end{tabular}
	\label{tab:topk1a}
\end{table*}

\begin{table*}[!ht]
	\centering	
	\caption{Comparison on Top-K recommendation with different combinations of blocks (\%). (Amazon-Book)}
	\vspace{-1em}
	\begin{tabular}{c|ccc|ccc}
		\hline
		method&F1@5&F1@10&F1@15&ndcg@5&ndcg@10&ndcg@15\\ \Xhline{1.5pt}
		SDRL&\bf{9.0582($\pm$0.07)}&\underline{11.4625($\pm$0.12)}&\underline{12.2647($\pm$0.14)}&\bf{13.7224($\pm$0.18)}&\underline{13.3115($\pm$0.14)}&\underline{14.8334($\pm$0.1)}\\
		SDRL'(int+ext)&\underline{8.8361($\pm$0.07)}&\underline{11.1924($\pm$0.1)}&\underline{12.0149($\pm$0.11)}&\underline{13.3619($\pm$0.17)}&\underline{12.9737($\pm$0.12)}&\underline{14.5178($\pm$0.08)}\\
		SDRL'(int+oth)&8.1967($\pm$0.1)&10.4707($\pm$0.12)&11.3059($\pm$0.14)&12.3478($\pm$0.18)&12.1008($\pm$0.1)&13.5747($\pm$0.08)\\
		SDRL'(ext+oth)&\underline{9.0518($\pm$0.08)}&\bf{11.4633($\pm$0.09)}&\bf{12.3021($\pm$0.1)}&\underline{13.7024($\pm$0.12)}&\bf{13.3441($\pm$0.09)}&\bf{14.9128($\pm$0.06)}\\
		SDRL'(whole)&1.908($\pm$0.05)&2.4953($\pm$0.05)&2.8446($\pm$0.03)&3.0402($\pm$0.11)&2.9335($\pm$0.05)&3.3497($\pm$0.02)\\
		\hline
	\end{tabular}
	\label{tab:topk2a}
\end{table*}

\begin{table}[ht]
	\centering
	\small
	\caption{Comparison on item classification with different combinations of blocks (\%). (ml-ls)}
	\vspace{-1em}
	\begin{tabular}{c|ccc}
		\hline
		method&recall&precison&micro\_F1\\ \Xhline{1.5pt}
		SDRL&\underline{59.7927}&\underline{71.2913}&\underline{65.0372}\\
		SDRL'(int+ext)&\bf{62.1688}&\bf{74.263}&\bf{67.6797}\\
		SDRL'(int+oth)&\underline{61.9033}&\underline{73.8443}&\underline{67.3484}\\
		SDRL'(ext+oth)&52.5572&62.046&56.908\\
		SDRL'(whole)&39.9923&42.0991&41.0177\\
		\hline
	\end{tabular}
	\label{tab:cla0a}
\end{table}

\begin{table}[ht]
	\centering
	\small
	\caption{Comparison on item classification with different combinations of blocks (\%). (ml-10m)}
	\vspace{-1em}
	\begin{tabular}{c|ccc}
		\hline
		method&recall&precison&micro\_F1\\ \Xhline{1.5pt}
		SDRL&\bf{80.0098}&\bf{80.0832}&\bf{80.0464}\\
		SDRL'(int+ext)&\underline{77.1831}&\underline{77.2037}&\underline{77.1933}\\
		SDRL'(int+oth)&\underline{76.558}&\underline{76.4345}&\underline{76.4961}\\
		SDRL'(ext+oth)&51.3015&49.2074&50.2283\\
		SDRL'(whole)&47.1581&43.9449&45.492\\
		\hline
	\end{tabular}
	\label{tab:cla1a}
\end{table}

\begin{table}[ht]
	\centering
	\small
	\caption{Comparison on item classification with different combinations of blocks (\%). (Amazon-Book)}
	\vspace{-1em}
	\begin{tabular}{c|ccc}
		\hline
		method&recall&precison&micro\_F1\\ \Xhline{1.5pt}
		SDRL&\underline{94.4971}&\underline{83.5058}&\underline{88.6618}\\
		SDRL'(int+ext)&\bf{98.2341}&\bf{86.0925}&\bf{91.7634}\\
		SDRL'(int+oth)&\underline{97.9884}&\underline{85.9162}&\underline{91.556}\\
		SDRL'(ext+oth)&53.6879&49.6705&51.6008\\
		SDRL'(whole)&48.922&45.5&47.1487\\
		\hline
	\end{tabular}
	\label{tab:cla2a}
\end{table}

\begin{figure*}
	\centering
	\subfigure[SDRL, ml-ls]{
		\includegraphics[width=0.3\linewidth]{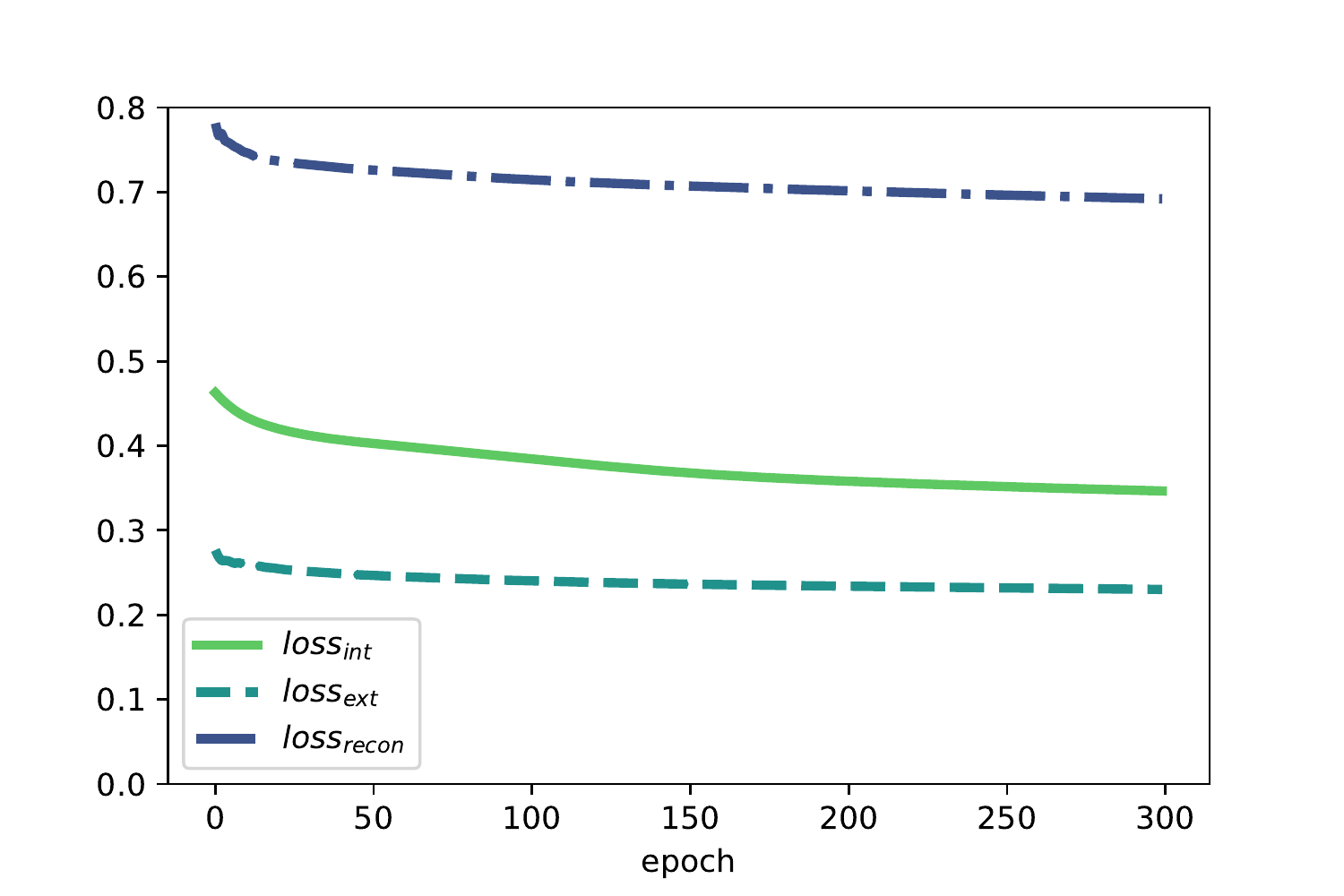} 
		\label{fig:loss0}
	}
	\subfigure[SDRL, ml-10m]{
		\includegraphics[width=0.3\linewidth]{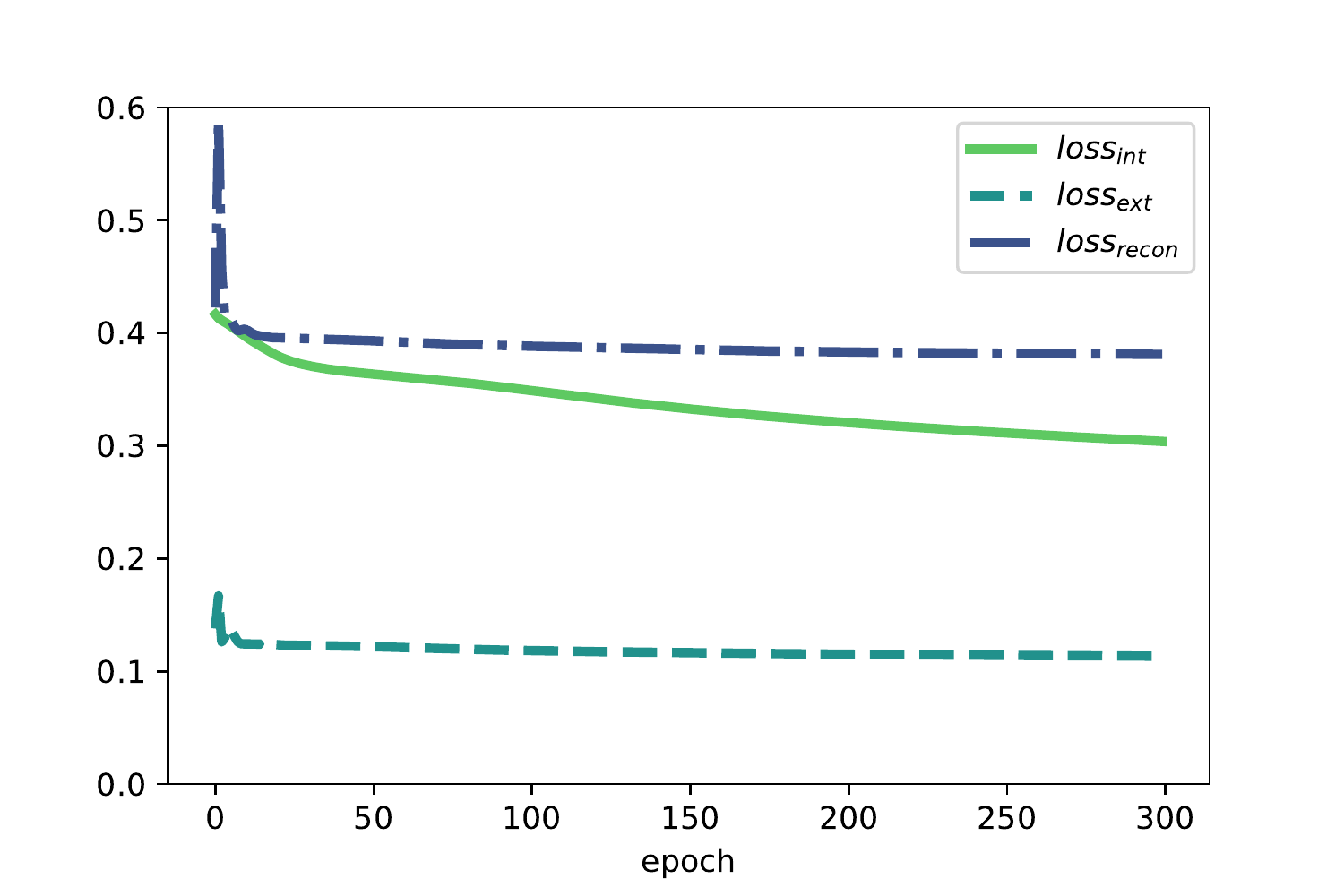} 
		\label{fig:loss1}
	}
	\subfigure[SDRL, Amazon-Book]{
		\includegraphics[width=0.3\linewidth]{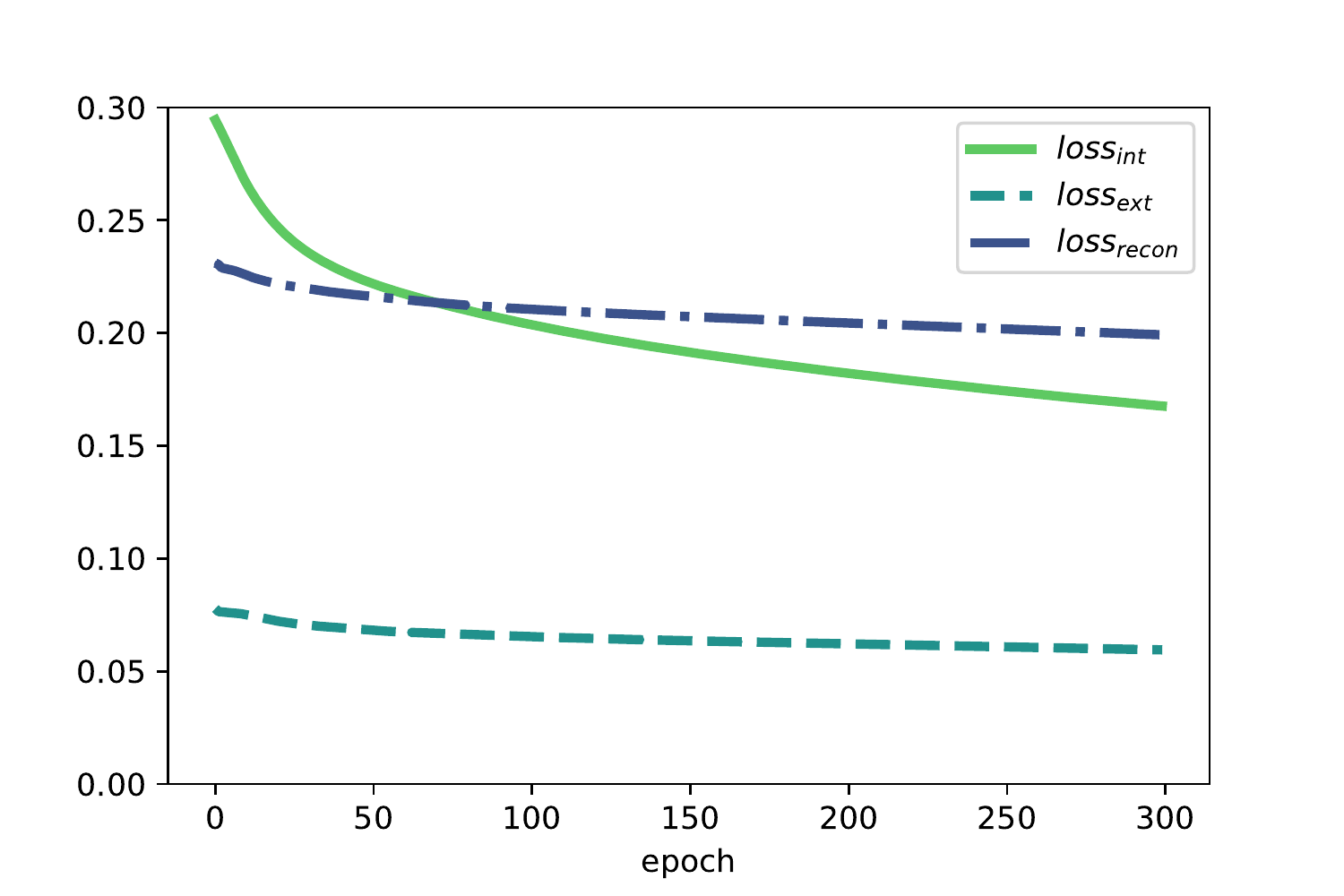} 
		\label{fig:loss2}
	}
	\subfigure[SDRL'(int+other), ml-ls]{
		\includegraphics[width=0.3\linewidth]{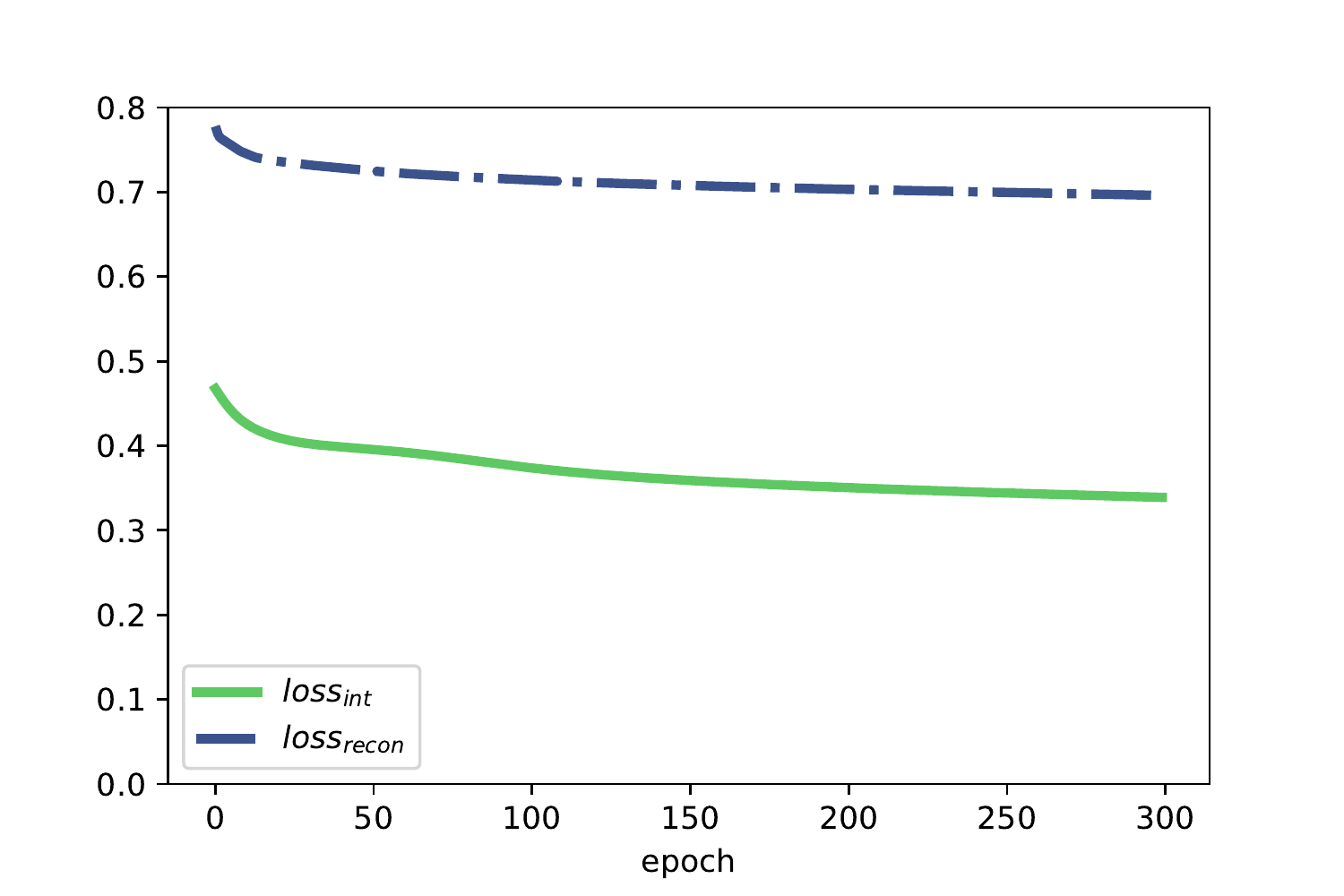} 
		\label{fig:loss30}
	}
	\subfigure[SDRL'(int+other), ml-10m]{
		\includegraphics[width=0.3\linewidth]{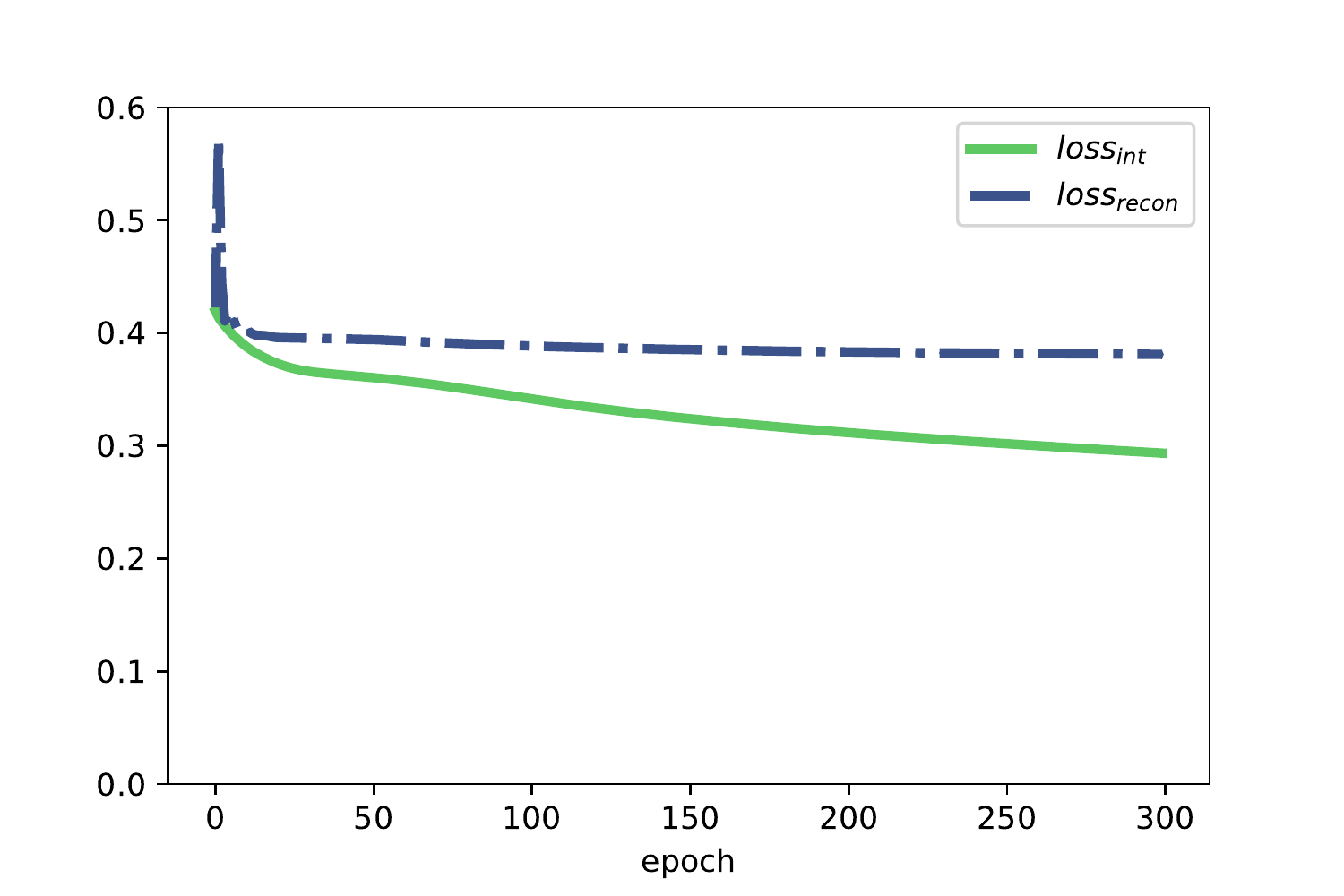} 
		\label{fig:loss31}
	}
	\subfigure[SDRL'(int+other), Amazon-Book]{
		\includegraphics[width=0.3\linewidth]{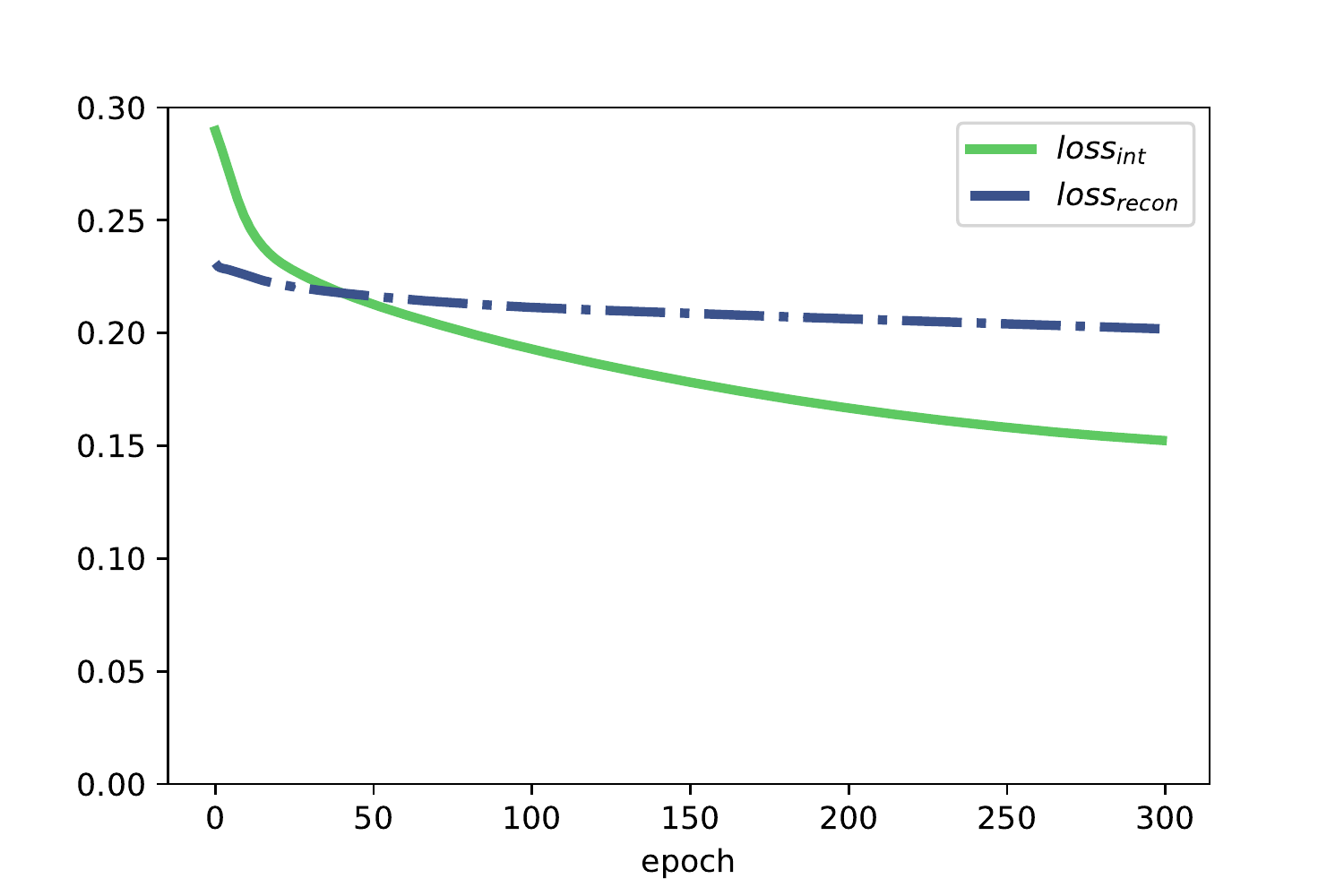} 
		\label{fig:loss32}
	}
	\subfigure[SDRL'(ext+other), ml-ls]{
		\includegraphics[width=0.3\linewidth]{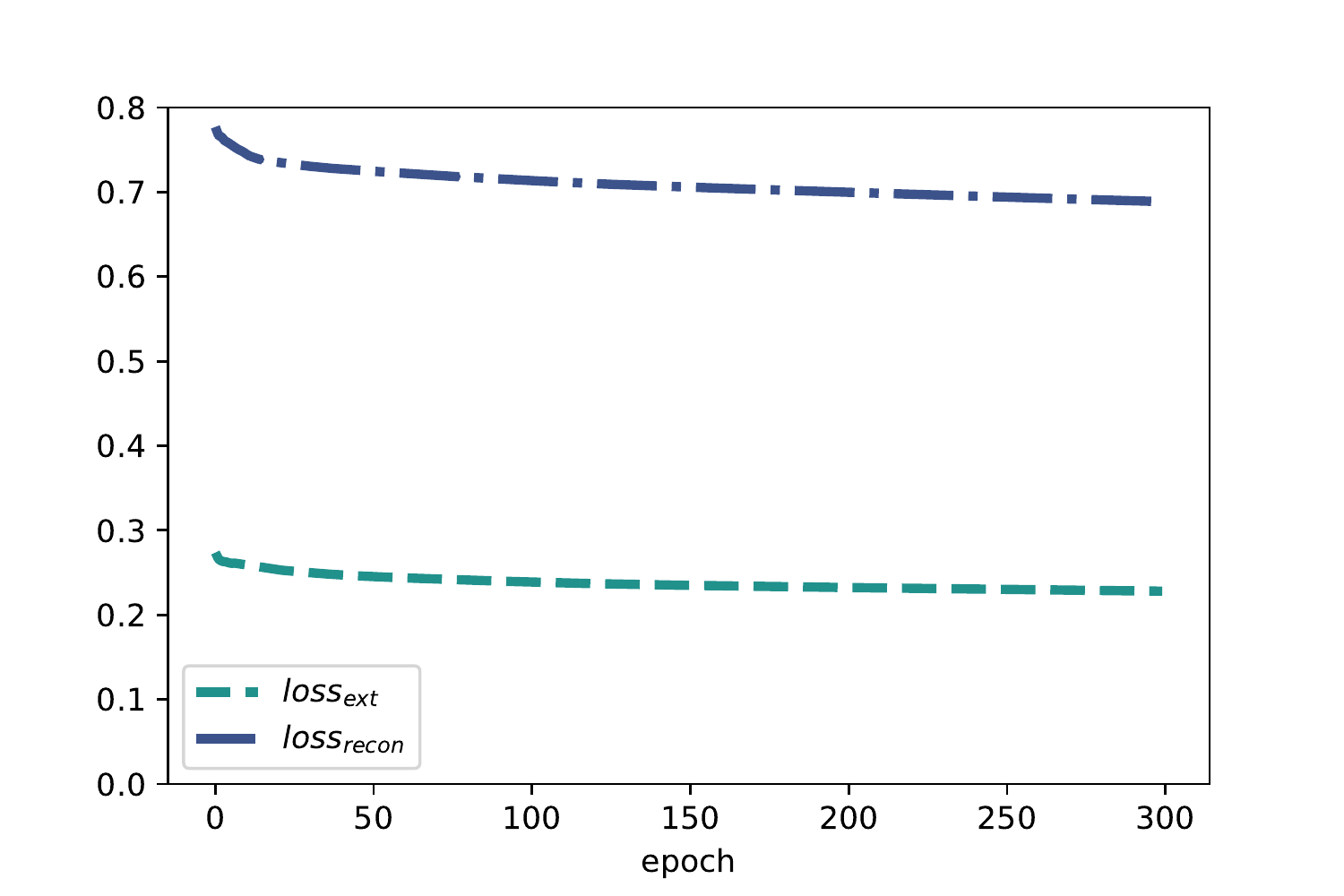} 
		\label{fig:loss40}
	}
	\subfigure[SDRL'(ext+other), ml-10m]{
		\includegraphics[width=0.3\linewidth]{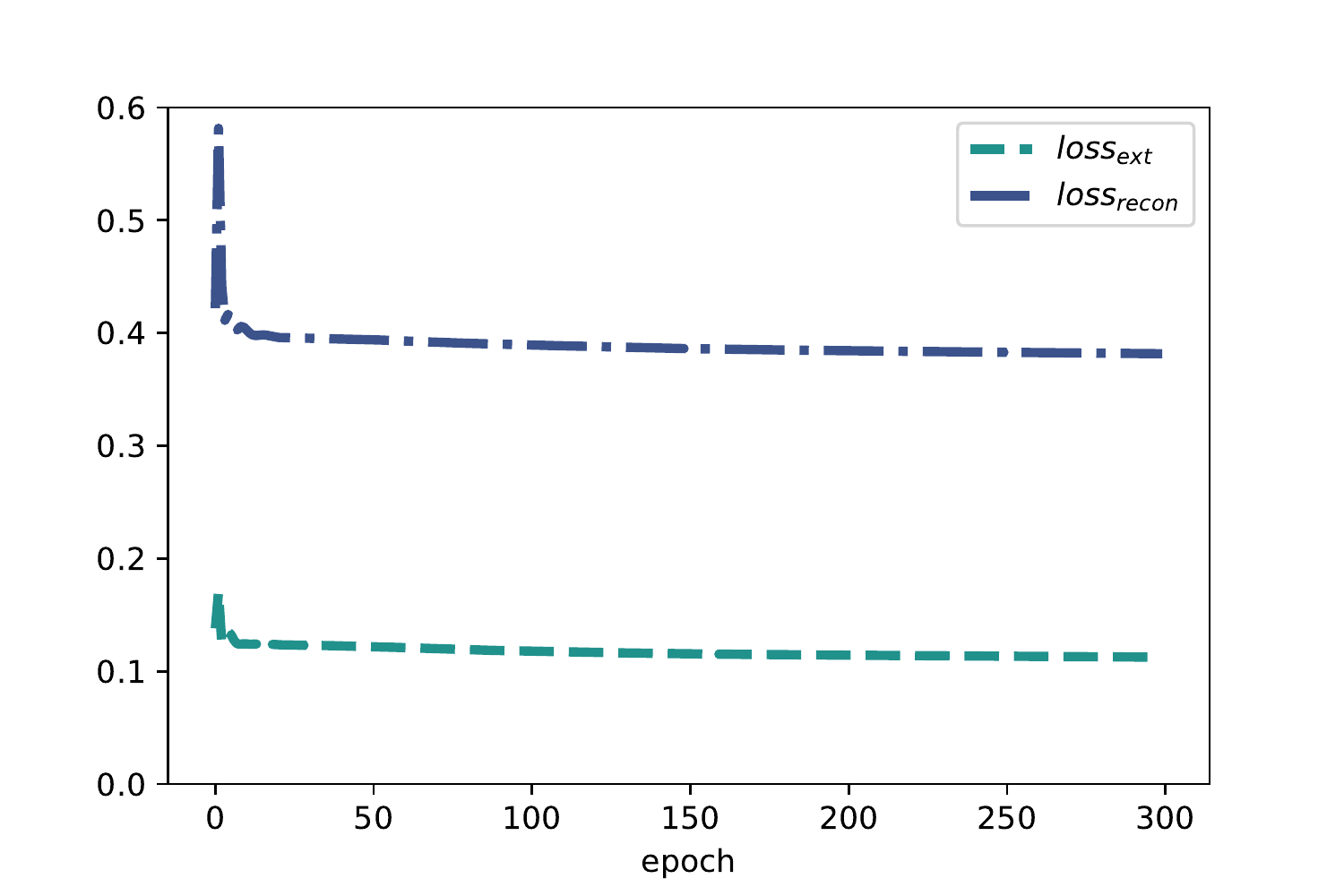} 
		\label{fig:loss41}
	}
	\subfigure[SDRL'(ext+other), Amazon-Book]{
		\includegraphics[width=0.3\linewidth]{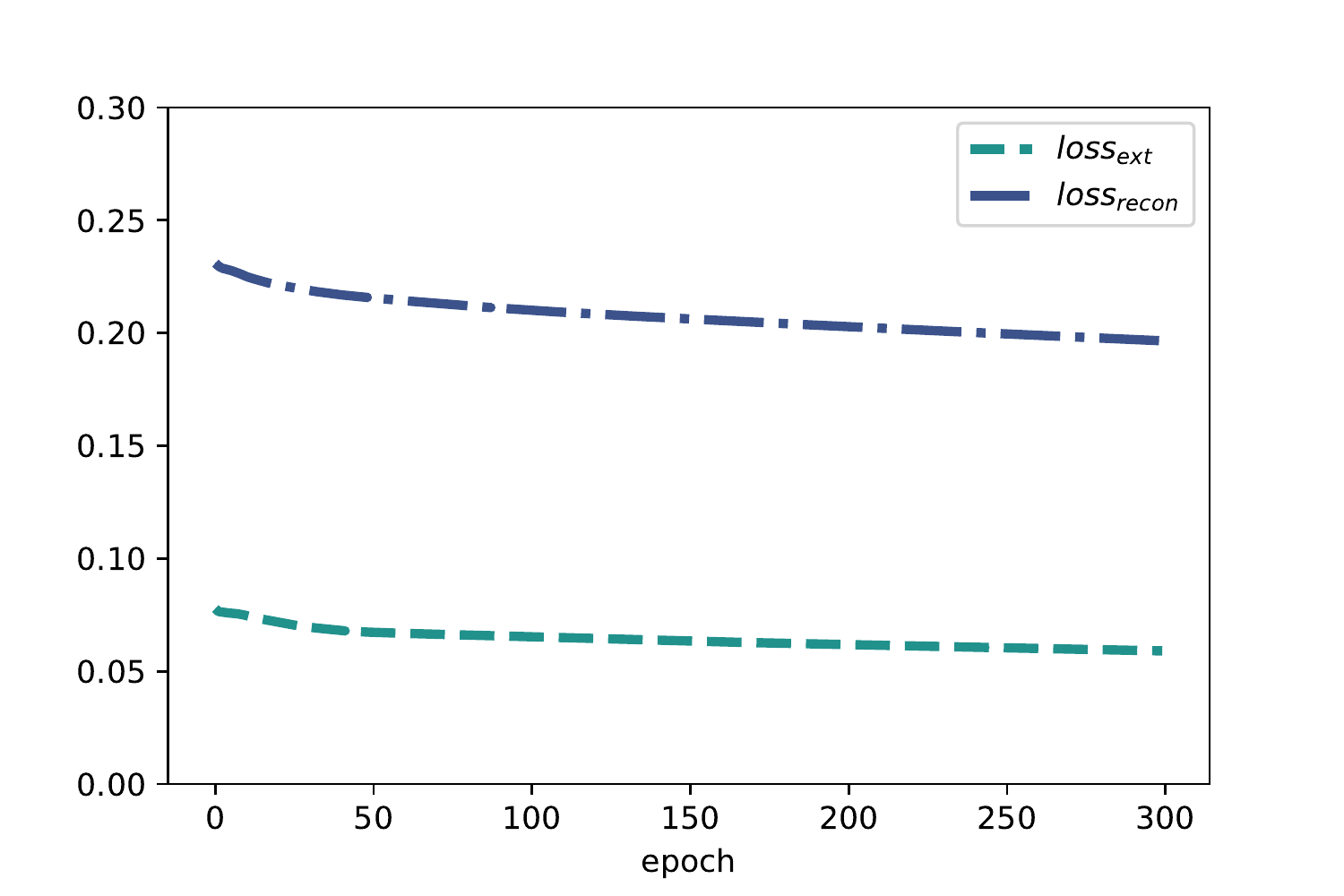} 
		\label{fig:loss42}
	}
	\caption{Illustration of Losses in the model training.}
	\label{fig:loss}
\end{figure*}

{\em Top-K Recommendation.} We generate Top-5, 10 and 15 items according to predicted ratings as recommendations. The comparison results on three datasets are shown in Tables \ref{tab:topk0}, \ref{tab:topk1} and \ref{tab:topk2}, respectively. Based on the observations of comparison results, we find that our method SDRL achieves steady improvements compared with baselines. It at least improves F1 (ndcg) by 20.41\% on MovieLens-latest-small, 35.24\% on MovieLens-10m and 0.29\% on Amazon-Book. It demonstrates that the proposed representation method SDRL significantly improves the Top-K recommendation performance especially on MovieLens datasets ml-ls and ml-10m. The possible reason is that more dense interactions (as shown in Table \ref{tab:dataset}) reflect richer features and it could promote (semi-)disentangled representation more accurate overall. 

{\em Item Classification.}
Item classification also plays an important role in recommendation system such as cold-start task \cite{zhao_categorical-attributes-based_2018}. Since VAECF and MacridVAE do not generate item representations, we only utilize NetMF and ProNE as  baselines and item categories as labels in item classification. We employ item embedding as the input and a MLPClassifier \footnote[2]{https://scikit-learn.org/} as the classification algorithm for all methods. The comparison results are shown in Tables \ref{tab:cla0}, \ref{tab:cla1} and \ref{tab:cla2}, respectively. We observe that our method SDRL outperforms baselines on three datasets with the help of category supervision.

\subsection{Ablation Analysis} \label{sec:ablation}
To study the impacts of various blocks in SDRL, we develop the ablation study based on the same setting as comparison experiment. Not only highlighting the best values in bold, we also underline the second and third best values in the results.

{\em Top-K Recommendation.}
Based on the observations of comparison results in Tables \ref{tab:topk0a}, \ref{tab:topk1a} and \ref{tab:topk2a}, we have three major findings. (1) Overall, {\em SDRL} consisting of three blocks and {\em SDRL'(int+ext)} achieve a relatively better performance. We infer that both category- and rating-based features play an important role in Top-K recommendation. (2) Another find is that variants with separated blocks usually outperforms that with the only one block {i.e., \em SDRL'(whole)}. It demonstrates that semi-disentanglement significantly improves the performance in Top-K recommendation.

(3) We also find a difference between the results on MovieLens datasets (i.e., ml-ls and ml-10m) and Amazon-Book. On MovieLens datasets, {\em SDRL'(int+oth)} outperforms {\em SDRL'(ext+oth)}, while on Amazon-Book {\em SDRL'(ext+oth)} shows a better performance. That's to say, the $internal\ block$ has a bigger impact on MovieLens datasets and the $external\ block$ takes more effects on Amazon-Book. We conduct further analysis and find that the difference may have a close relation with the statistical characteristics of datasets. As is shown in Fig. \ref{fig:sta_gen}, on MovieLens datasets, over 80\% users relate with at least 14 categories (77.78\%); and on Amazon-Book, over 80\% users only relate with at least 12 categories (54.55\%). The statistical characteristics of datasets could affect the role of different features in downstream tasks, and (semi-)disentanglement increases the flexibility to emphasize certain features.

We also conduct a deeper analysis in terms of model optimization, to explore what causes the difference. As introduced in Section \ref{sec:method}, we employ $loss_{int}$ (in Equation \ref{eq:int}) to encourage the $internal\ block$ to express relatively more category-based features. Meanwhile, the $external\ block$ (in Equation \ref{eq:ext})  is expected to contain more interactive features with the optimization of $loss_{ext}$. In addition, $loss_{recon}$ (in Equation \ref{eq:recon})  optimizes the whole reconstruction of users, items and ratings, which would make the $internal\ block$ also contain some interactive features. 

We draw the trends of each parts of loss functions for three methods (i.e., SDRL, {\em SDRL'(int+oth)} and {\em SDRL'(ext+oth)}) respectively in Fig. \ref{fig:loss}. We observe that there is a similar trend among $loss_{recon}$, $loss_{int}$ and $loss_{ext}$ on MovieLens datasets in Fig.s \ref{fig:loss0}, \ref{fig:loss30} and \ref{fig:loss40} (or in Fig.s \ref{fig:loss1}, \ref{fig:loss31} and \ref{fig:loss41}). That is, the training process fairly optimizes the $internal\ block$, $external\ block$ and the whole reconstruction. However, on Amazon-Book in Fig.s \ref{fig:loss2}, \ref{fig:loss32} and \ref{fig:loss42}, $loss_{int}$ obviously decreases faster than the other two losses. 

It demonstrates that in the model training on Amazon-Book, the optimization for $internal\ block$ plays a more important role. The bias on Amazon-Book pushes the embeddings to contain more category-based features and relatively fewer rating-based features. The effect appears especially significant in {\em SDRL'(int+oth)}, in which $loss_{ext}$ is removed from loss function. However, in Top-K recommendation, interactive features based on ratings may play a major part. It probably explain why {\em SDRL'(int+oth)} performs poorer than {\em SDRL'(ext+oth)} on Amazon-Book. Similarly, on MovieLens datasets, {\em SDRL'(ext+oth)} generates representations without the supervision of category-based information. Meanwhile, for {\em SDRL'(int+oth)}, the fair optimization would encourage embeddings contain both category information and interactive features. Therefore, {\em SDRL'(int+oth)} outperforms {\em SDRL'(ext+oth)} on MovieLens datasets.

Last but not least, the difference also validates the importance of employing multiple features as supervision for different blocks in the representation learning, i.e., supervised (semi-)disentangled representation.

\subsection{Hyper-parameter Analysis} \label{sec:para}
\begin{figure}[h]
	\centering
	\includegraphics[width=0.8\linewidth] {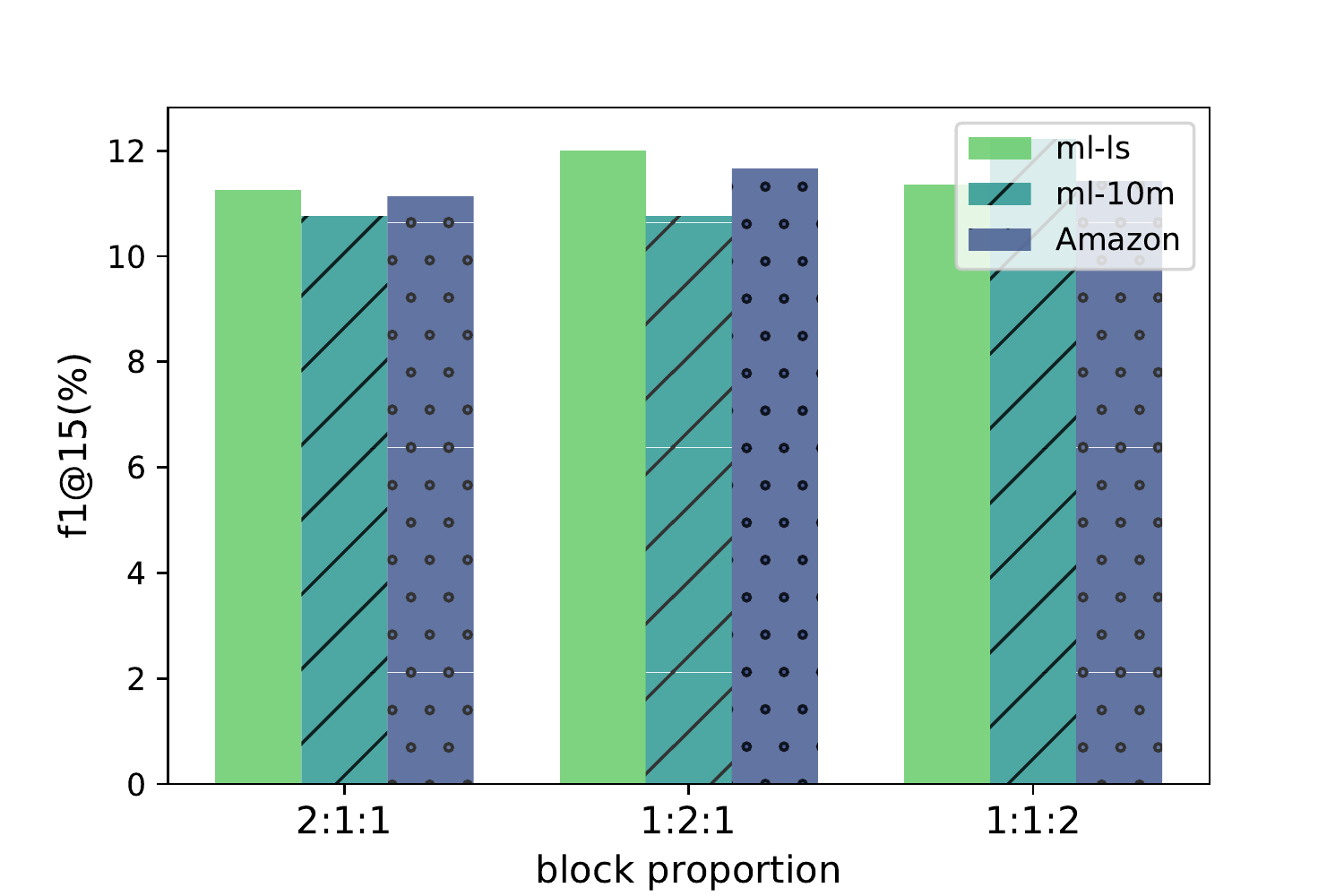}
	\vspace{-1em}
	\caption{Impact of Block Proportion.}
	\label{fig:hyper}
\end{figure}

It is flexible to adjust the proportion of $internal\ block$, $external\ block$ and $other\ block$ in the embeddings. We set the proportion as 2:1:1, 1:2:1 and 1:1:2 to test their performance on Top-15 recommendation. The results in Fig. \ref{fig:hyper} show the variant with proportion 2:1:1 (i.e., a bigger proportion for $internal\ block$) outperforms the others on MovieLens datasets. On Amazon-Book, the variant with a bigger proportion for $external\ block$ performs better. Therefore, we set the proportion in SDRL as  2:1:1 on the Movielens datasets and 1:2:1 on Amazon-Book.

\subsection{Representation Visualization} \label{sec:vis}
In SDRL, $internal\ block$ and $external\ block$ are expected to respectively express features from individual itself and user-item interactions. We visualize node representation to qualitatively analyze the semi-disentanglement of two parts. 

\subsubsection{Visualization based on Item Representation} 
\begin{figure}[h]
	\centering
	\subfigure[ml-ls (int)]{
		\includegraphics[width=0.47\linewidth]{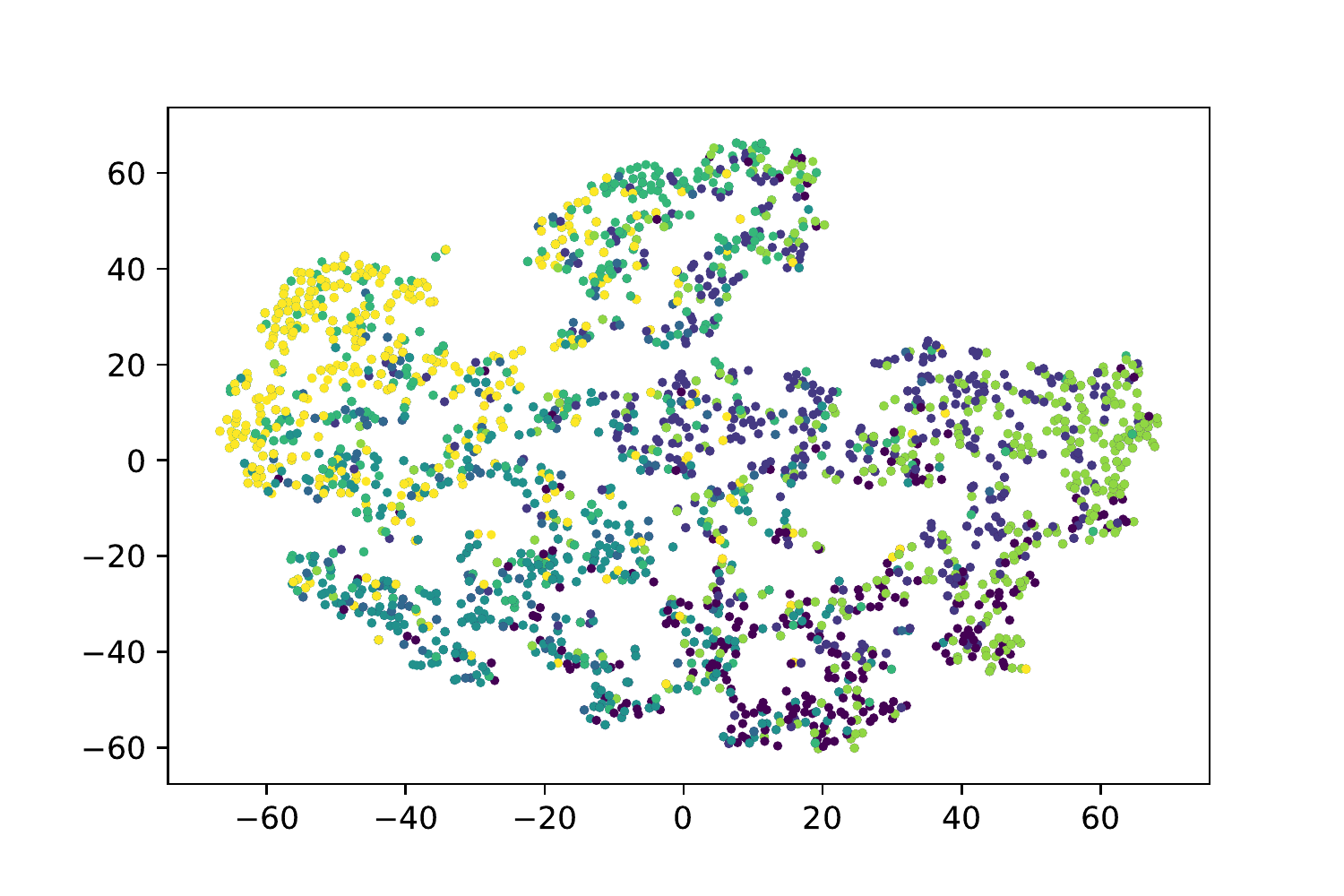} 
	}
	\subfigure[ml-ls (ext)]{
		\includegraphics[width=0.47\linewidth]{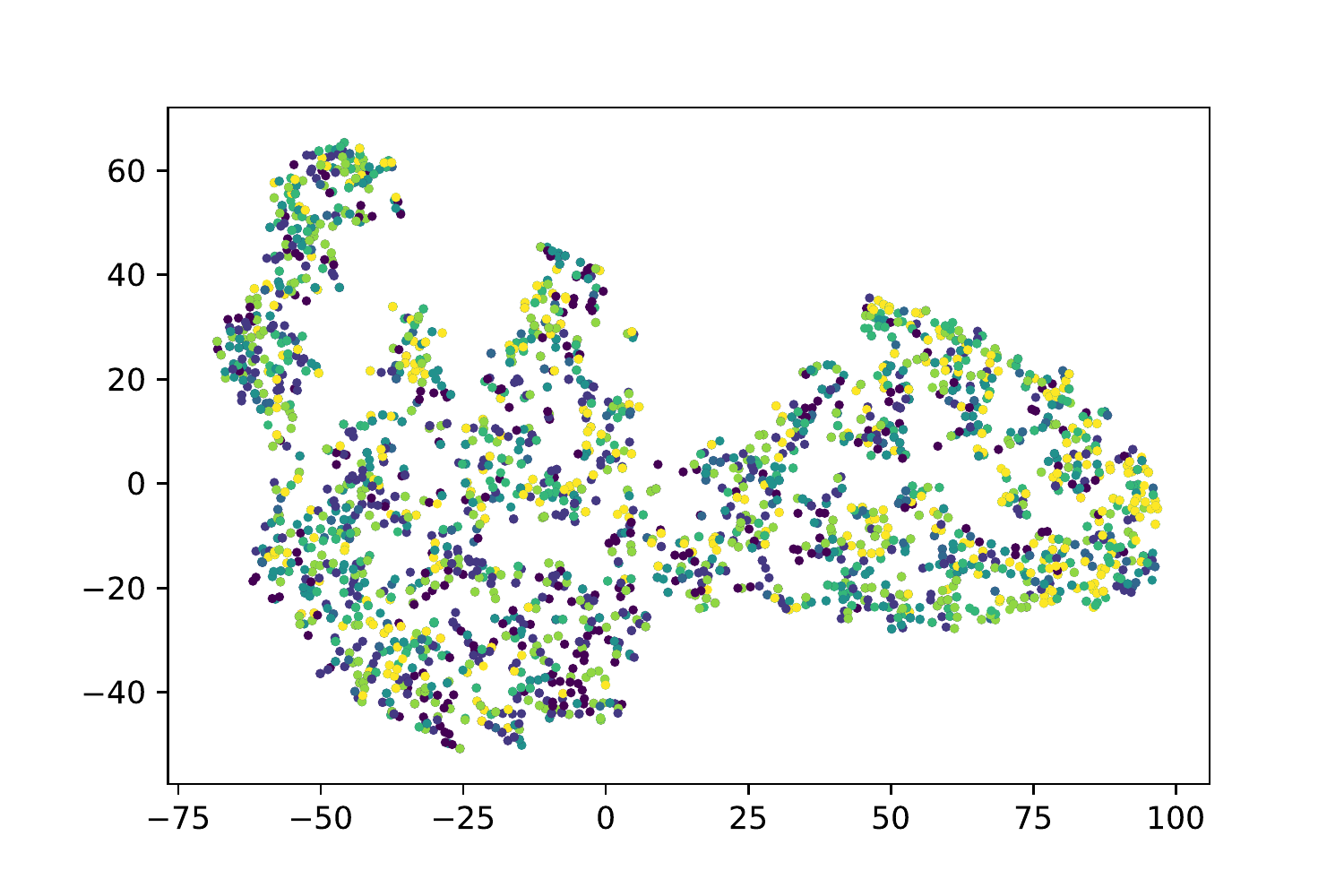} 
	}
	\subfigure[ml-10m (int)]{
		\includegraphics[width=0.47\linewidth]{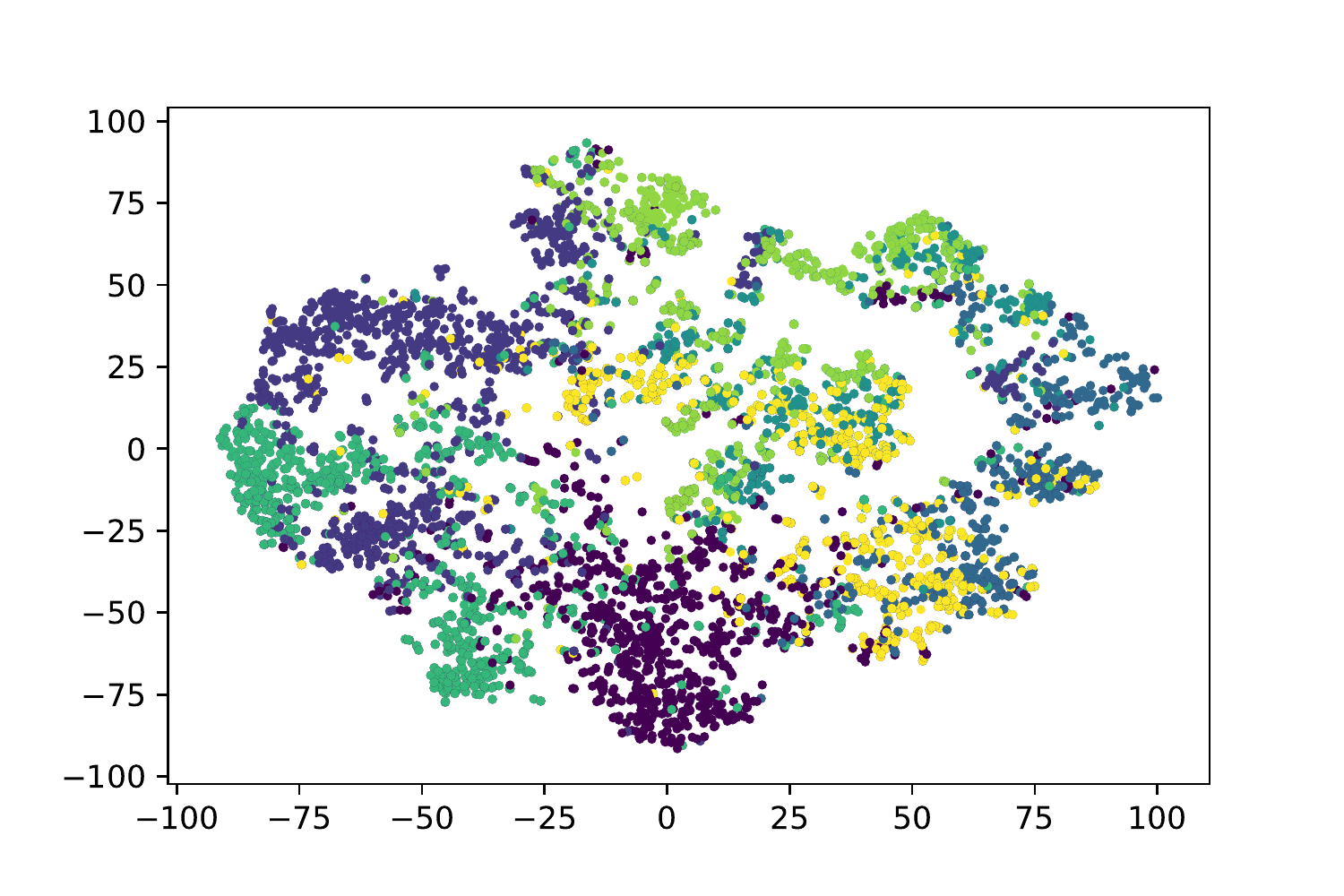} 
	}
	\subfigure[ml-10m (ext)]{
		\includegraphics[width=0.47\linewidth]{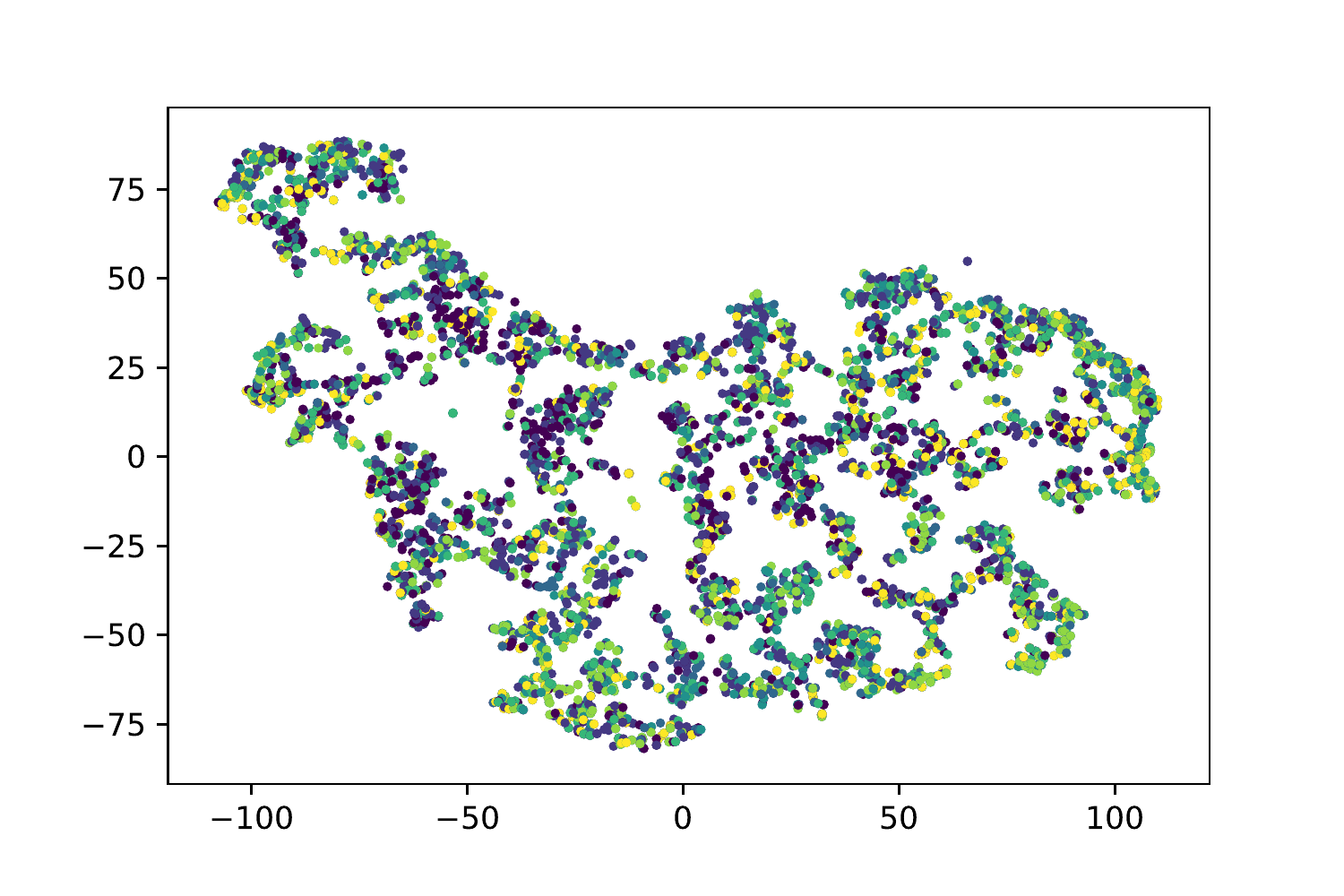} 
	}
	\subfigure[Amazon-Book (int)]{
		\includegraphics[width=0.47\linewidth]{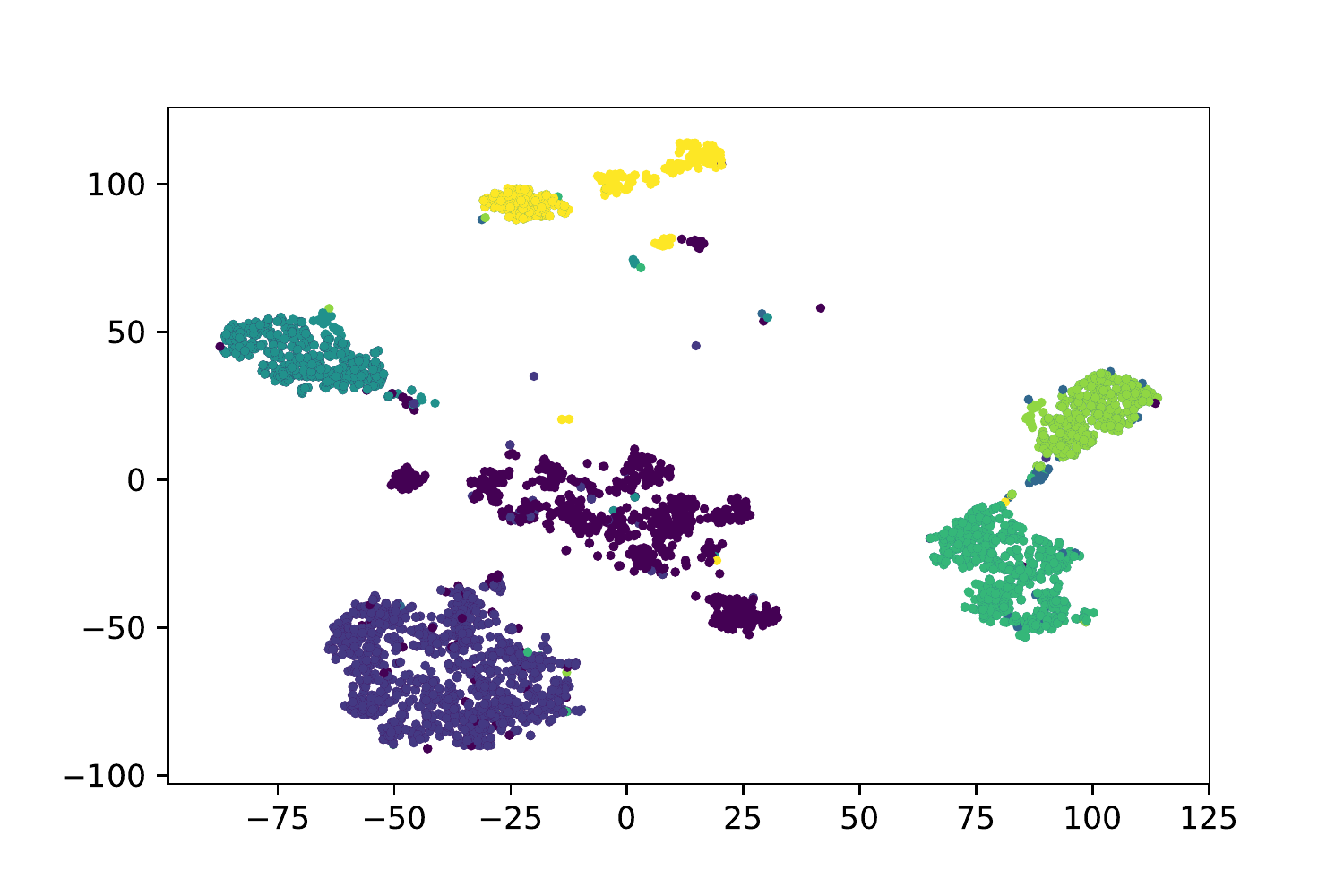} 
	}
	\subfigure[Amazon-Book (ext)]{
		\includegraphics[width=0.47\linewidth]{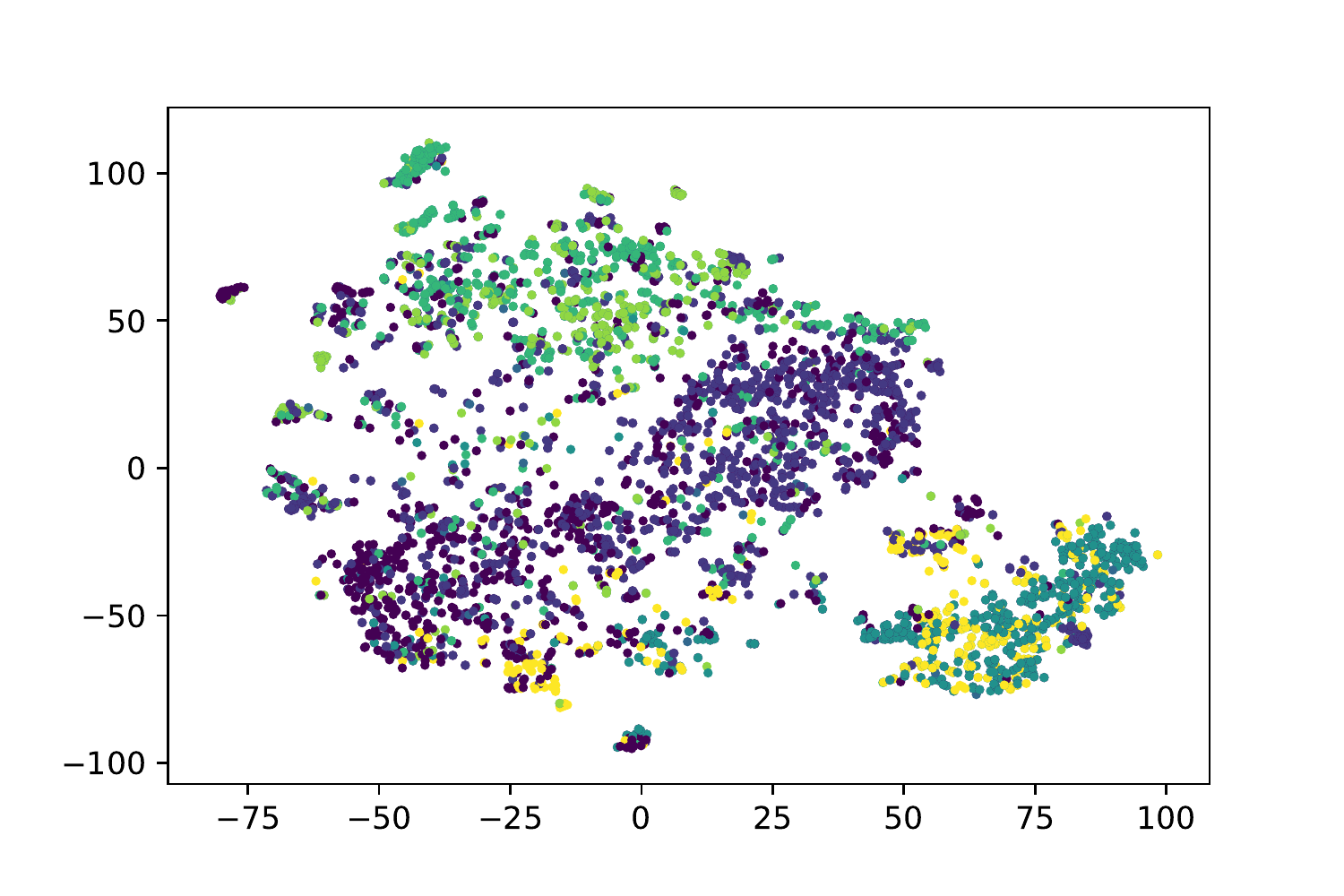} 
	}
	\vspace{-1em}
	\caption{Visualization of Item Representation.}
	\label{fig:vis}
\end{figure}

For clarity, we choose the $7$  most common categories as labels and visualize  $internal\ block$ and $external\ block$ of items using t-SNE \cite{rauber_visualizing_2016} respectively. The results are revealed in Fig. \ref{fig:vis}. We could find that in the left figures, nodes in same color (i.e., items belonged to the same category) are more clustered. It indicates that $internal\ block$ contains more category-related information than $external\ block$, which is consistent with our expectation.

\subsubsection{Visualization based on User Representation}
\begin{figure}[h]
	\centering
	\subfigure[Visualization of User Representation.]{
		\includegraphics[width=\linewidth]{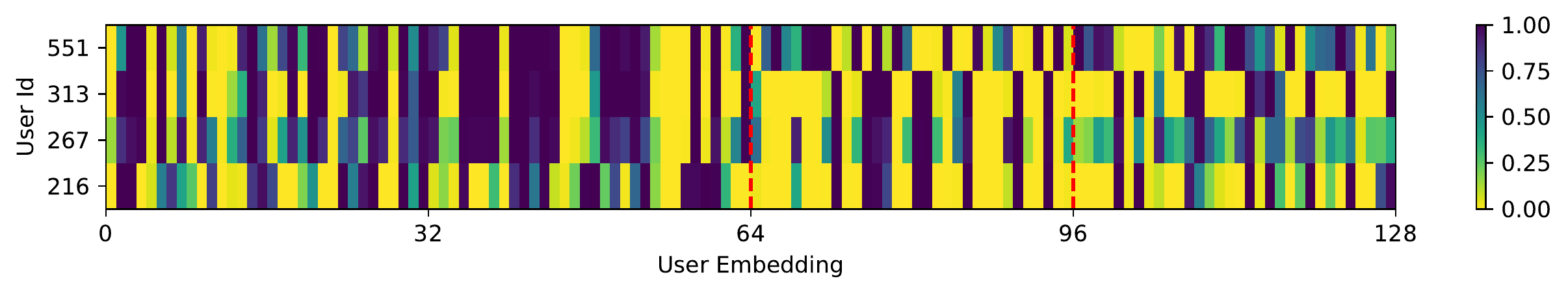} 
	}
	\subfigure[Visualization of User Preference on Category.]{
		\includegraphics[width=0.56\linewidth]{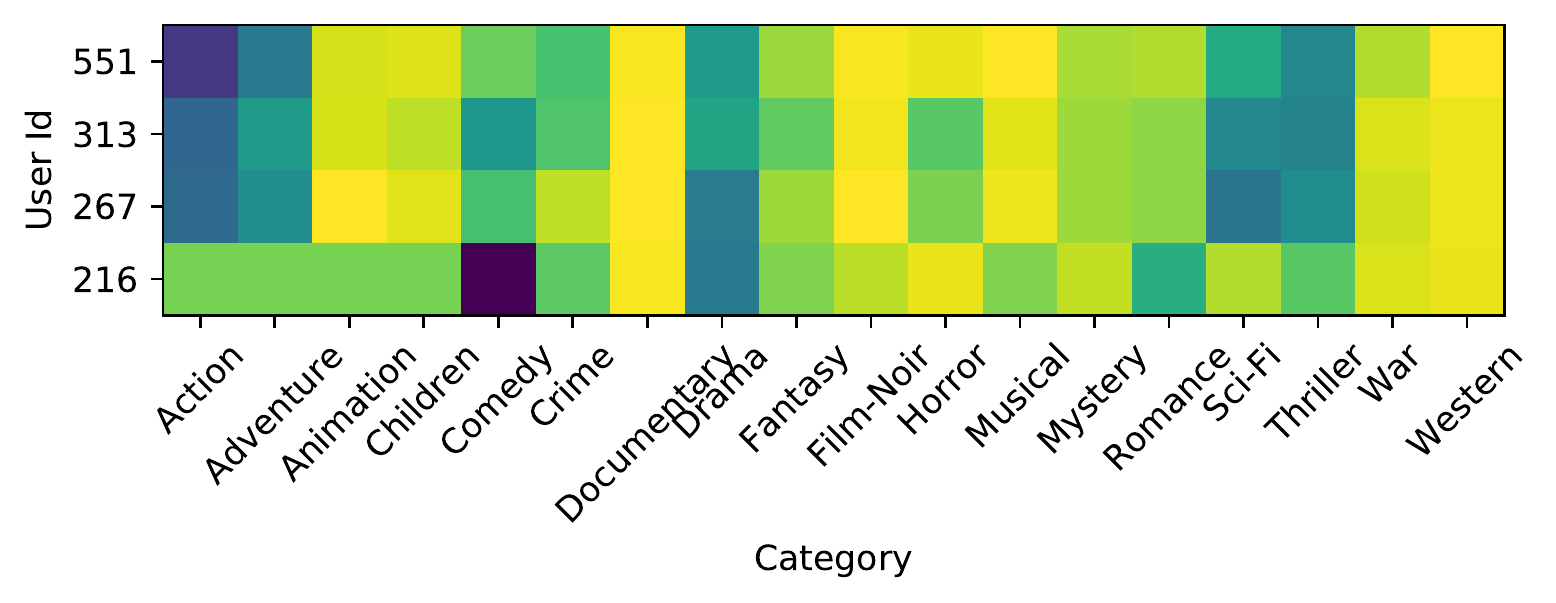}
		\label{fig:vis_user_b} 
	}
	\subfigure[Statistics of Common Items.]{
		\includegraphics[width=0.39\linewidth]{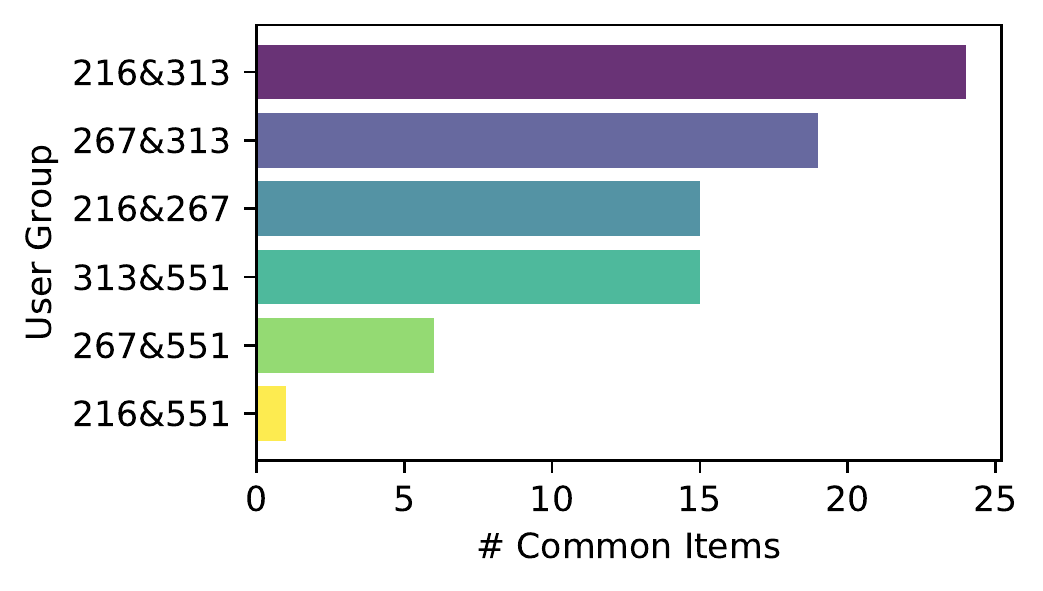} 
		\label{fig:vis_user_c}
	}
	\vspace{-1em}
	\caption{Relation among User Representation, Preference on Category and Common Items.}
	\label{fig:vis_user}
\end{figure}

We also select four users (whose user IDs are 551, 267, 313 and 216) on MovieLens-latest-small to visualize their representation in Fig. \ref{fig:vis_user}. For 128-dimension embeddings on MovieLens-latest-small, the first 64 bits represent $internal\ block$, the next 32 bits $external\ block$ and the last 32 bits $other\ block$. Among them, users No.551, No.267 and No.313 share a similar $internal\ block$ representation while the embedding of user No. 216 shows a different one. Based on source data displayed in Fig. \ref{fig:vis_user_b}, we find that the former three users take interest in action, adventure and thriller movies while the later likes comedy movies the most. Meanwhile, there is similar information on $external\ block$ among users No.216, No.267 and No.313 (especially users No.216 and No.313), while user No.551 is a little bit different. Based on the statistics of common items in Fig. \ref{fig:vis_user_c}, we could find that there are relatively more common items among users No.216, No. 313 and No.267 than user No.551 and the others.

Shortly, in line with the expectation, $internal\ block$ and $external\ block$ express more features of category and user-item interactions, respectively. It also improves the interpretability of node representation as well as offer clues to generate explanations. For instance, when the recommended item better matches the target user on $internal\ block$, an explanation based on category maybe more reasonable, while that on $external\ block$ indicates interaction-based explanation is more suitable.

\subsection{Summary of Experiments}
In summary, we have the following findings in the experiments. (1) Overall, SDRL gains stable and general improvements, demonstrating its effectiveness in representation learning. (2) In consistent with expectation, $internal\ block$ and $external\ block$ respectively express more category-based and interactive features. (3) Semi-disentanglement enhances the explainability and generality in terms of representation in recommendation system. 

\section{Related Work}
We would briefly review related works on representation in recommendation system and disentangled representation.

\subsection{Representation in Recommendation System}
Representation learning transforms the sparse data in recommendation system into structured embeddings, providing great convenience for complex network process \cite{cui_survey_2019}. Based on the type of source data, existing representation methods could be divided into three categories: structure-based \cite{perozzi_deepwalk_2014,grover_node2vec_2016}, content-based \cite{zhang_prone_2019,zhao_deep_2020} and both-based \cite{wang_reinforced_2020,he_lightgcn_2020,fu_magnn_2020}. Among them, both-based approaches receive lots of attentions in recent years, e.g., graph neural network (GNN). It initializes representation with content features and then update it with structure information.

Our method SDRL also employs content and structure features as the input. The major difference between SDRL and GNN-based approaches is that SDRL embeds content- and structure-based information into different blocks (i.e., $internal\ block$ and $external\ block$ respectively) while they represent it as a whole.

\subsection{Disentangled Representation}
Disentangled representation learning aims to separate embedding into distinct and explainable factors in an unsupervised way \cite{bengio_representation_2013,do_theory_2020}. It has been successfully applied into computer vision \cite{nemeth_adversarial_2020}. Recently, Locatello et al. \cite{locatello_challenging_2019,locatello_commentary_2020} demonstrate that unsupervised disentangled representation learning without inductive biases is theoretically impossible. To deal with this problem, some works with (semi-)supervision are proposed \cite{locatello_disentangling_2020,chen_weakly_2020}. However, the assumption of factor independence makes typical disentangled representation not applicable for recommendation.  

Our method differs from existing disentangled representation works in that we do not separate the whole embedding (i.e., we preserve $other\ block$ for no specific meanings) and we do not explicitly force the independence of different factors. In this way, we could preserve more complicated relations in recommendation system, while achieving proper disentanglement.

\section{Conclusions}
To improve both disentanglement and accuracy of representation, we propose a semi-disentangled representation method for recommendation SDRL. We take advantages of category features and user-item ratings as the supervision for the proposed $internal\ block$ and $external\ block$. To the best of our knowledge, it is the first attempt to develop semi-disentangled representation as well as improve disentanglement with supervision in recommendation system. We conduct intensive experiments on three real-world datasets and the results validate the effectiveness and explainability of SDRL. In the future work, we would try to extract more labels and utilize them to separate the explainable part into fine-grained features for extensive applications. We are also interested to deeply study the unexplainable part to manage the uncertainty. 
\bibliography{ref.bib}
\bibliographystyle{plain}
\end{document}